# On the Variability of Critical Size for Homogeneous Nucleation in a Solid-State Diffusional Transformation

Pooja Rani[†], R.M. Raghavendra[†] and Anandh Subramaniam[*]

*Materials Science and Engineering, Indian Institute of Technology Kanpur,*

*Kanpur-208016, INDIA, Ph.: +91(512)259 7215*

[†] *Authors with equal contribution and affiliation reflects the place of origin of the work*

[*]*Author for correspondence:* anandh@iitk.ac.in

In a solid-state diffusional phase transformation involving nucleation and growth, the size of the critical nucleus for a homogeneous process ($r^*_{homo} = r^*$) has been assumed to be a time invariant constant of the transformation. The strain associated with the process has a positive energy contribution and leads to an increase in the value of $r^*$, with respect to that for nucleation from a liquid. With the progress of such a transformation, the strain energy stored in the matrix increases and nuclei forming at a later stage encounter a strained matrix. Using devitrification of a bulk metallic glass as a model system, we demonstrate that $r^*$ is *not* a cardinal time invariant constant for homogeneous nucleation and can increase or decrease depending on the strain energy penalty. We show that the assumption regarding the constancy of $r^*$ is true only in the early stages of the transformation and establish that the progress of the transformation leads to an altered magnitude of r*, which is a function of the microstructural details, geometrical variables and physical parameters. With the aid of high-resolution lattice fringe imaging and computations of $r^*$, we further argue that, 'liquid-like' homogeneous nucleation can occur and that the conclusions are applicable to a broad set of solid-state diffusional transformations. The above effect 'opens up' a lower barrier transformation pathway arising purely from the internal variables of the system.

A first order diffusional phase transformation proceeds by nucleation and growth. The control of nucleation and growth is of paramount importance in diverse scientific and engineering applications [1]. In the ambit of the classical nucleation theory (CNT) [2,3], the nucleation step occurs 'uphill' in the Gibbs free energy (G) and embryos below a critical size ($r^*$) tend to revert to the parent phase [4]. In an undercooled system held at constant pressure and temperature, the nucleation step involves a 'random statistical fluctuation' in the metastable phase, which leads to the formation a product nucleus of size $r^*$ [5]. The net Gibbs free energy change for the formation of a crystal from a solid ($\Delta G_{homo}$) via homogeneous nucleation is given by:

$$\Delta G_{\text{homo}} = -(\Delta G_{\text{V}} \cdot V) + (\gamma \cdot A) + V \cdot \Delta G_{\text{S}} \qquad (1)$$

where, $\Delta G_{\text{V}}$ is the Gibbs volume free energy difference between the parent and product phases, '$V$' is the volume of the product phase, '$A$' is the interfacial area, $\gamma$ is the interfacial energy and $\Delta G_{\text{S}}$ is the strain energy associated with the transformation. For a nucleus of spherical shape of radius '$r$', equation (1) can be written as:

$$\Delta G_{\text{homo}} = -\tfrac{4}{3}\pi r^3 (\Delta G_{\text{V}} - \Delta G_{\text{S}}) + (4\pi r^2) \cdot \gamma \qquad (2)$$

The critical size for homogeneous nucleation ($r^*_{\text{homo}} = r^*$) is determined by finding the extremum of the function $\Delta G_{\text{homo}}(r)$. Two special cases of the nucleation of a crystal are noteworthy: (i) that from a liquid and (ii) that from an amorphous matrix. In the former case the strain energy term is absent and the critical size of the nucleus is: $r^*_{\text{liquid}} = 2\gamma / \Delta G_{\text{V}}$. The later case forms a model system of a solid to solid transformation, especially from an experimental perspective, due to the following attributes. (i) The degree of transformation is amenable to good control (via time and temperature). (ii) In the absence of 'interference' of the lattice fringes from the matrix, the crystallite size can be measured with relative ease, using the high-resolution lattice fringe imaging (HRLFI) technique. (iii) The isotropic properties of the matrix enable the simplification of the analysis. (iv) Sites of heterogeneous nucleation (like grain boundaries, dislocations, etc.) are absent, which can dominate over the homogenous nucleation processes in a typical solid to solid diffusional phase transformation.

For the specific case of the formation of a crystalline nucleus in an amorphous matrix, the strain energy associated with the transformation can be written as [6]:

$$E_{\text{Strain}} = \frac{8\pi r^3 f_{\text{m}}^2 G}{3} \frac{(1+\nu)}{(1-\nu)} \qquad (3)$$

where, '$f_{\text{m}}$' is the linear mismatch between the nucleus and the glass matrix (the linear mismatch is one third of the volumetric mismatch), '$G$' & '$\nu$' are shear modulus and Poisson's ratio respectively of the product phase, which is assumed to be isotropic in this simplified formula. Equation (3) contains only the elastic properties of the product phase, while in reality the strain energy is partitioned between the matrix and the product phases and hence should involve properties of both phases [7]. The inclusion of the strain energy term leads to an increase in the magnitude of $r^*$, which is given by:

$$r^*_{bulk} = \frac{2\gamma}{(\Delta G_V - \Delta G_S)} = r^*_\infty \qquad (4)$$

Where, $\Delta G_S = \frac{E_{Strain}}{V}$ is the misfit strain energy per unit volume. Here, the '∞' in the subscript represents to a single nucleus in an infinite matrix.

Needless to point out, the shape of the nucleus will determine the magnitude of the energies involved in equation (4) and this variable has been investigated in literature [8,9]. Lee et al. [8] has highlighted the role of the strain energy in determining the shape of the critical nucleus. Zhang et al. [9] have computed the morphology of the critical nucleus taking into account the elastic anisotropy and have further shown that the shape of nucleus can be plates, needles or cuboids. In a related context, the effect of shear strain on the shape of the nucleus has also been investigated [10].

Experimentally, HRLFI has been the preferred technique for the study of nucleation; especially in the determination of the critical nucleus size [11,12]. In a recent fascinating work, Zhou et al. [13] have used atomic electron tomography to study nucleation in a model FePt system. Computational techniques like molecular dynamics and density functional theory have also been used to study the nucleation behaviour in solids [14-16]. The critical size can be altered due to the presence of preferred sites of nucleation (heterogeneous nucleation) [17,18] or in the presence of defects and their associated strain fields, which leads to a reduced strain energy penalty [19,20]. In many circumstances, a combination of these effects may be superposed [21].

The effect of an altered strain energy penalty on the magnitude of $r^*$, has been investigated for the case of martensitic transformations [5,22]; wherein the parent matrix may develop considerable strain with the progress of the transformation. The scenario wherein $r^*$ is altered due to nucleation in nanoscale volumes has also been investigated [23]. In this work, a good match was observed between the value of $r^*$ computed using a combination of CNT and finite element computations; with that experimentally measured using HRLFI. This work also served to validate the utility of CNT in the context of computation of $r^*$ for the devitrification of a bulk metallic glass (BMG). In an interesting study, Gómez et al [24] showed the variability of $r^*$ due to the effect of curvature of the substrate on the nucleation of two-dimensional phases. Shen et al. [25] have incorporated the effect of the stress state of the matrix, arising from the pre-existing microstructure, on nucleation. They have further incorporated the same in a phase field model. The effect of atom trapping on the critical nucleus size has been investigated by Erdélyi et al. [26]. In a related context, the effect of strain energy on the formation of an interfacial melt has also investigated [27]. An important point which emerges from the survey of the literature is that, the critical radius for homogeneous nucleation ($r^*$) has been considered as cardinal time invariant constant for a given diffusional transformation.

In the context of nucleation in a solid-to-solid diffusional transformation, the standard analysis involves the formation of a single nucleus in an infinite matrix. With the progress of the transformation this picture is altered and nuclei forming at a later stage do so in the proximity of pre-existing grown crystallites. The strain field of these pre-existing crystallites not only interacts amongst themselves in a complex fashion; but also alters the 'energy accounting', which arises due to the altered strain energy penalty for nucleation at a specific coordinate.

In the current work, we use high-resolution lattice fringe imaging (HRLFI) and finite element computations to achieve the following tasks. (1) Establish that $r^*$ is not a cardinal time invariant constant for a given solid to solid phase transformation. (2) Compute the value of $r^*$ for model scenarios arising during the progress of transformation. (3) Elucidate that $r^*$ is function of the microstructural details, geometrical variables and physical parameters like anisotropy. (4) Argue that 'liquid-like' nucleation can occur with the progress of the transformation. The formation of $Cu_{10}Zr_7$ crystal (Cmca, oC68, a = 12.68 Å, b = 9.31 Å, c = 9.35 Å [28]) from an amorphous matrix (composition of $(Cu_{64}Zr_{36})_{96}Al_4$) is considered as a model system to achieve the abovementioned tasks.

We use the CNT in our computations and aspects regarding its utility are discussed later in the script and in the supplementary information. The following parameters are required for the computation of the value of $r^*$ (using equation (4): (i) $\Delta G_V$, (ii) crystal-glass interfacial energy ($\gamma$), (ii) $E_{strain}$. Further, the elastic constants of the phases and the value of the misfit strain ($f_m$) are required for the determination of $E_{strain}$. We consider here briefly the methodology for the calculation of these quantities and details can be found elsewhere ([23], supplementary information). For the glass to crystal transformation, the value of $\Delta G_V$, reported by Zhou and Napolitano [29], is used for the calculation. The value of $\gamma$ is determined using a formula derived using the negentropic model [30,31] ($\gamma = 0.03$ J/m$^2$). The strain energy is computed using a finite element model, based on the Eshelby formalism [32].

As alluded to before, with the progress of the transformation the stress state in the matrix becomes complex due to a overlap of the stress fields arising from individual crystallites. Model three-dimensional finite element simulations have been used for the computation of the strain energy of the system, to elucidate the effect of nucleation in the vicinity of single and multiple crystallites. Figure 1 shows a schematic of a finite element model used for the calculation of the strain energy of a nucleus in the proximity of a larger crystallite. Eigenstrains are imposed in selected regions (Regions-A) of the numerical model (Figure 1a), corresponding to the volume misfit between the crystal and the glass. Anisotropic elastic properties have been used for the crystal. The matrix being glass, has isotropic material properties. The interaction of the strain field depends on the relative crystallographic orientation between the crystal and the nucleus, which is varied in a discrete manner. To compute the strain energy of a single nucleus in an infinite matrix, eigenstrains are imposed in Region-A only. The corresponding value of $r^*$ is labeled as $r^*_{bulk}$. A 3D view of the system is shown in Figure 1b. In the model considered the strain energy penalty for nucleation is a function of the: (i) shape of the nuclei, (ii) relative crystallographic orientation between the large crystal and the nucleus, (iii) distance of the nucleus from the crystal and (iv) size of the large crystal. The effect of these variables has been studied for specific value of the parameters. The relative crystallographic orientation between the 'large crystal' (L) and the nucleus (N) is designated by specifying two parallel basis vectors of the orthogonal set ($[U_1V_1W_1]_L \| [u_1v_1w_1]_N$ & $[U_2V_2W_2]_L \| [u_2v_2w_2]_N$).

A plot of the hydrostatic stress contours ($\sigma_{hyd}$) in the presence of a spherical nucleus (r = 1.1 nm) at a distance of 2 nm (*d*) from the grown crystal (R = 5 nm) is shown in Figure 1c. The relative orientation of the crystallites is: $[\bar{1}\bar{1}2]_L \| [010]_N$ & $[\bar{1}10]_L \| [100]_N$ ($[110]_L \| [001]_N$). From the stress plot it is seen that tensile and compressive regions exist in the matrix around the large/grown crystallite. In the figure the nucleus is positioned in a compressive region of the large crystal. If nucleation occurs in the tensile region of the large crystallite, this will lead to an increase in the magnitude of $r^*$. The converse will be true for the nucleation in the compressive region of the large crystallite. The details regarding the finite element models and the importance of the role of hydrostatic stress and other variables in the process, is described in the supplementary material.

*(Insert Figure 1 about here)*

Figure 2 shows the plot of the variation of critical nucleation size ($r^*$) as a function of the distance *(d)* from the grown crystallite. The value of the strain energy, which is an input in the computation of $r^*$, is computed using the model in Figure 1a. The plots correspond to two relative orientations of the crystallites: (O$_1$) $[\bar{1}\bar{1}2]_L \| [010]_N$ and (O$_2$) $[010]_L \| [010]_N$. It is seen that $r^*$ can increase or decrease depending on the tensile or compressive nature of the stress state in the matrix (schematically illustrated in Figure 1b). As expected, the magnitude of the influence of the large crystal increases with a decreasing magnitude of 'd'. The plots (Figure 2a) display an asymmetry between the two orientations, which arises due to the inherent elastic anisotropy of the crystal.

The plot of $\Delta G_{homo}$ versus 'r' is shown for following three cases are shown as insets to the figure. (i) Nucleation in the absence of a large crystal (Figure 2c). In this case $r^* = r^*_{bulk}$. (ii) Nucleation in the compressive region corresponding to orientation $O_1$ and d = 4 nm (Figure 2d). (iii) Nucleation in the tensile region corresponding to orientation $O_2$ and d = 4 nm (Figure 2e).

*(Insert Figure 2 about here)*

A brief outline of the experimental details is considered here and further details can be found in the supplementary information. The $(Cu_{64}Zr_{36})_{96}Al_4$ BMG is produced by arc melting followed by suction casting in a copper mold. The following heat treatments have been followed to obtain partially crystallized samples: (P1) 200°C for 120 min, (P2) 325°C for 30 min and (P3) 200°C for 120 min. The temperatures for the heat treatment are based on the nucleation and growth rate curves (supplementary material). In P1 the sample is annealed in the nucleation dominated regime to obtain a sparse dispersion of nuclei in an amorphous matrix. In P2 the nuclei are grown in the growth dominated regime to obtain crystallites in the size range of 5-15 nm. In P3, akin to P1, the treatment is carried out in the nucleation dominated regime to obtain nuclei between the larger crystallites. The above treatments are based on the following rationale: the nucleation is P1 is in a strain free matrix, with little interference between strain fields of the nuclei; while the nucleation in P3 is in the strain field of pre-existing crystallites. These crystallites have grown during the P2 treatment. A total of seven samples were investigated to obtain HRLFI from nuclei. A point to be noted here is that, BMGs in contrast to marginal glasses are produced by a slow cooling rate.

Figure 3 shows a compilation of experimental (HRLFI) and computational results. The computed values of $r^*$ are for diverse configurations, which are shown schematically as insets to the figure. The abscissa (ΔE) is a measure of the strain energy cost difference between nucleation in a strain-free matrix, versus that in a strained matrix. The different configurations considered give rise to a range of ΔE values and a negative value of ΔE implies a reduced strain energy cost for nucleation. Additional insets to the figure show selected HRLFI from nuclei, which have a range of $r^*$ values. The limit of the LFI technique to measure crystallite sizes is about 0.55 nm, which corresponds to five bright fringes and is marked as a broad band in the figure. The details regarding this aspect and the computational models can be found in the supplemental material. The following salient observations can be made from a study of Figure 3.

(A-i) (a) The nuclei which formed after the treatment P1 (i.e., nucleation in an unstrained matrix) have $r^*$ values of 1.42±0.03. This value matches well with the computed value of $r^*_{bulk}$ ($r^*_{bulk}(Computed) = 1.4\ nm$). The small scatter in the experimentally measured nuclei sizes ($r^*_{bulk}(Exp)$) is to be noted.

(b) The experimentally measured $r^*$ values for samples with P3 treatment span a range: from that exceeding $r^*_{bulk}$, to that close to the value of $r^*_{liquid}$.

(c) The number of nuclei with $r^* > r^*_{bulk}$ is much less than for the number of nuclei with $r^* < r^*_{bulk}$.

(A-ii) Nucleation in the tensile region(s) of the large crystal(s) leads to an increase in value of $r^*$ and the reverse is true for nucleation in the compressive region(s).

(A-iii) The presence of multiple 'favourably oriented' crystals accentuates the effect on $r^*$.

A-i (a) implies that $r^*$ has a constant value in the absence of stress in the matrix. In a transformation, this scenario occurs at the early stages; wherein the density of nuclei is sparse and no growth has taken place. This result, additionally serves to validate the computational methodology adopted to determine the value of $r^*$. A-i (b) implies $r^*$ is not a cardinal constant for a solid to solid diffusional transformation and that *liquid like* nucleation can occur in a bulk specimen. This scenario of variable $r^*$ will prevail for most parts of the transformation. A-i (c) is a natural consequence of the fact that, presented with a choice between higher a lower barrier height, the system chooses the later. In other words, since crystalline order arises in an amorphous matrix via a 'statistical fluctuation', the probability of occurrence of smaller crystal exceeds that of a larger crystal. A-ii and A-iii highlight the fact that local microstructural details dictate the increase or decrease in the value of $r^*$ and that a significant decrease in the magnitude of $r^*$ can occur in a highly strained matrix. It is noteworthy that, the computational models with elastic anisotropy, are able to capture the essential aspects of the experimental results.

*(Insert Figure 3 about here)*

The results obtained in the current work are for a specific system (Cu-Zr-Al system); however, given that the material properties of this system are in the typical range, we expect a broader applicability of the conclusions (i.e., for solid to solid phase transformation, including that for the nucleation of a crystal from a crystalline matrix). A detailed discussion regarding this aspect can be found in the supplementary information and an explicit validation of the same forms scope for the future work. In the current work a reasonably good explanation of the experimental observations is made via a computational methodology involving various assumptions. An analysis regarding the same can be found in the supplemental material. This includes alternatives to the CNT [13,33] and alternate computational methods. A definitive proof, establishing that the glass originally formed is free of quenched-in nuclei (i.e. pre-existing crystallites), can be found in the supplementary information.

The following conclusions can be drawn from the current work, with respect to nucleation in a solid to solid diffusional phase transformation.

(1) With the progress of the transformation the strain energy penalty for the formation of a nucleus is altered, leading to a variability in the magnitude of the critical nucleus size ($r^*$) (i.e. $r^*$ is not a cardinal invariant).
(2) This 'time dependent' penalty is a function of the microstructural details, geometrical variables and physical parameters like anisotropy and crystallographic orientation.
(3) The interaction of stress fields of the nucleus with a prior formed crystal can lead to a decrease or an increase in the value of $r^*$. The magnitude of the depression can be significant, which can result in a *'liquid-like'* nucleation in a solid state transformation.
(4) The aforementioned effect 'opens up' a lower barrier transformation pathway arising purely from the internal variables of the system.

**Acknowledgments**

We thank Jai Singh Advanced Imaging Centre, I.I.T. Kanpur for providing the facility to do TEM. Dr. Raghavendra Tewari, Bhabha Atomic Research Centre, Mumbai and Prof. Madhav Ranganathan, I.I.T. Kanpur, are kindly acknowledged for the discussions. The financial assistance from Science and Engineering Research Board, Department of Science and Technology (SERB-DST), India under Grant No. EMR/2016/004987 is acknowledged.

**Competing interests**

The authors declare no competing interests.

**Author contributions**

Pooja Rani carried out the experiments. R.M. Raghavendra performed the computations. Anandh Subramaniam conceived the problem and worked out a solution strategy and further penned the manuscript.

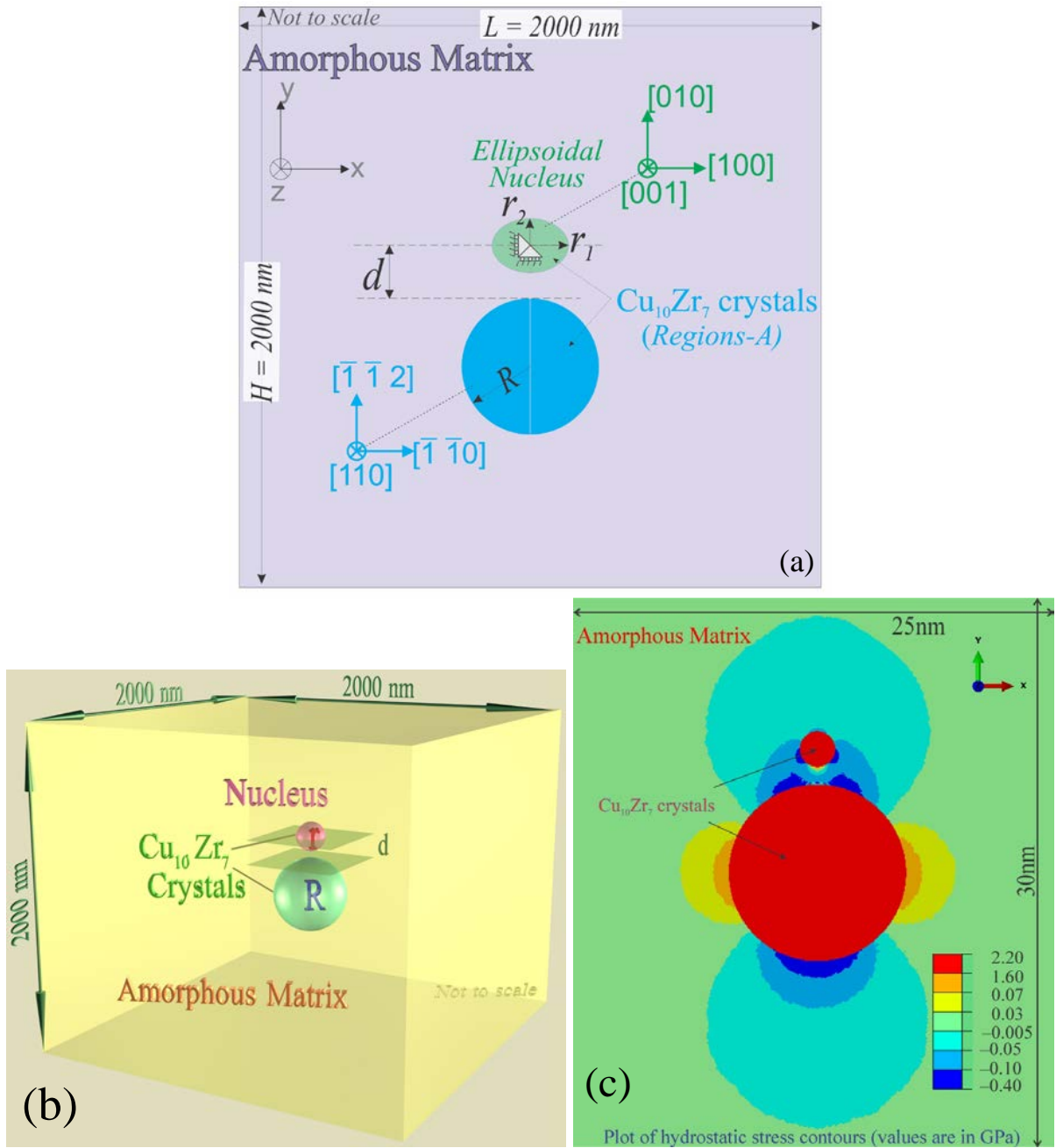

Figure 1. (a) A schematic of the finite element model (2D section of the 3D model) used to simulate the stress state of an ellipsoidal nucleus (with radii $r_1$ and $r_2$) in the proximity of a pre-existing crystal (of radius R). Eigenstrains are imposed in the Regions-A, corresponding to the volume mismatch between the $Cu_{10}Zr_7$ crystal and the amorphous matrix. The crystallographic orientation of the large/grown crystal is varied in a discrete manner to obtain specific orientation relationships. To simulate the case of a single nucleus in an 'infinite' matrix, eigenstrains are imposed only in Region-A. (b) A 3D view of the system. (c) A plot of the hydrostatic stress contours for the specific case of a spherical nucleus (r = 1.1 nm) at a distance (d) of 2 nm from the grown crystal. The radius of the large crystal (R) is 5 nm. The relative orientation of the crystallites marked in the figure corresponds to this special case and a zoomed-in view is depicted in the figure. The set (x, y, z) form the global coordinate set.

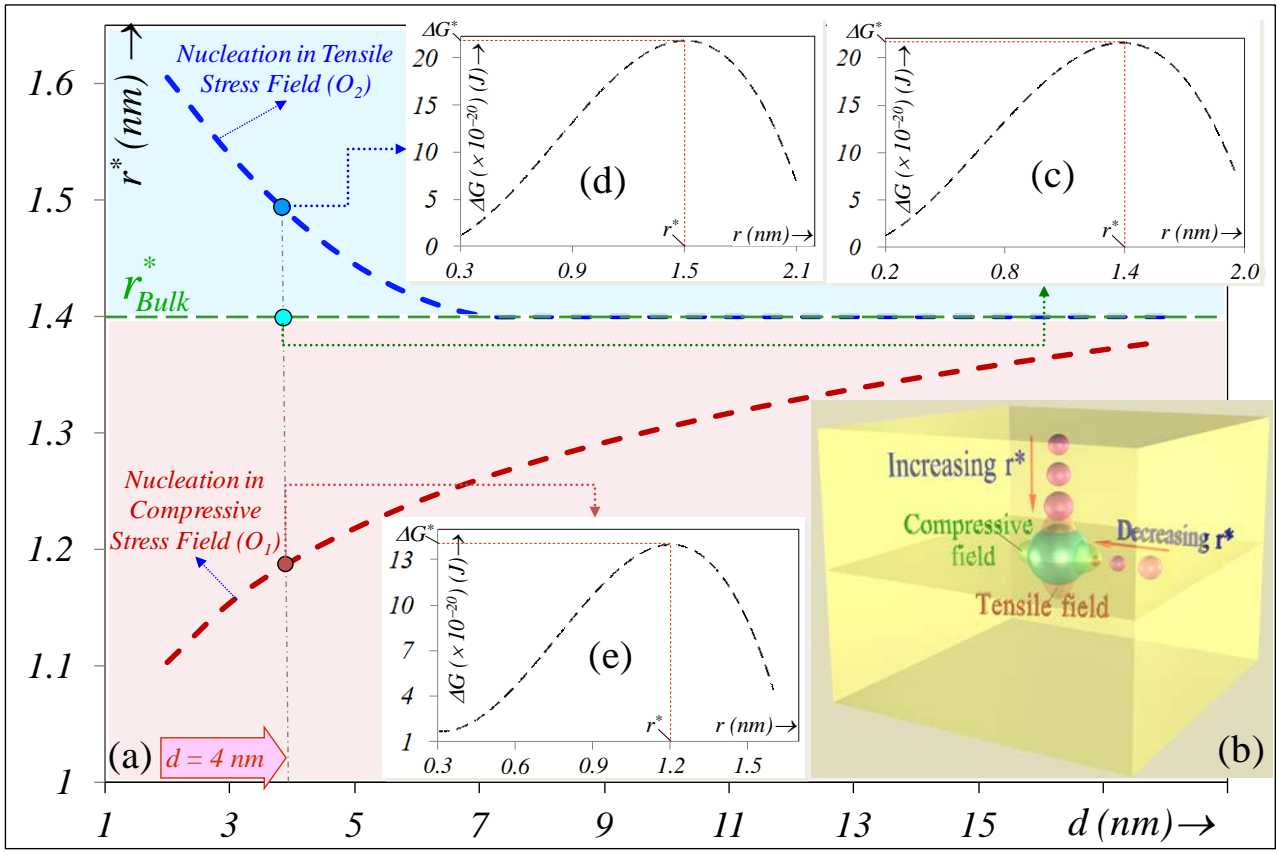

Figure 2. (a) Plot of the variation of critical nucleation size $r^*$ as a function of the distance 'd' from the grown crystallite, along two crystallographic orientations ((O1) $[\bar{1}\bar{1}2]_L \| [010]_N$ and (O2) $[010]_L \| [010]_N$). The horizontal dashed line corresponds to $r^*_{bulk}$ (i.e. nucleation in the absence of the large crystal). (b) Schematic illustrating the effect of the tensile and compressive regions of the large crystal in the magnitude of $r^*$. (c-e) Plot of the variation of $\Delta G_{homo}$ ($=\Delta G$) with '$r$' for three cases. Nucleation in the absence of the large crystal (c). Nucleation in the compressive region (d) and nucleation in the tensile region (e). In (d) and (e), d = 4 nm. The critical values of $\Delta G$ ($\Delta G^*$) and '$r$' ($r^*$) are marked in the figures.

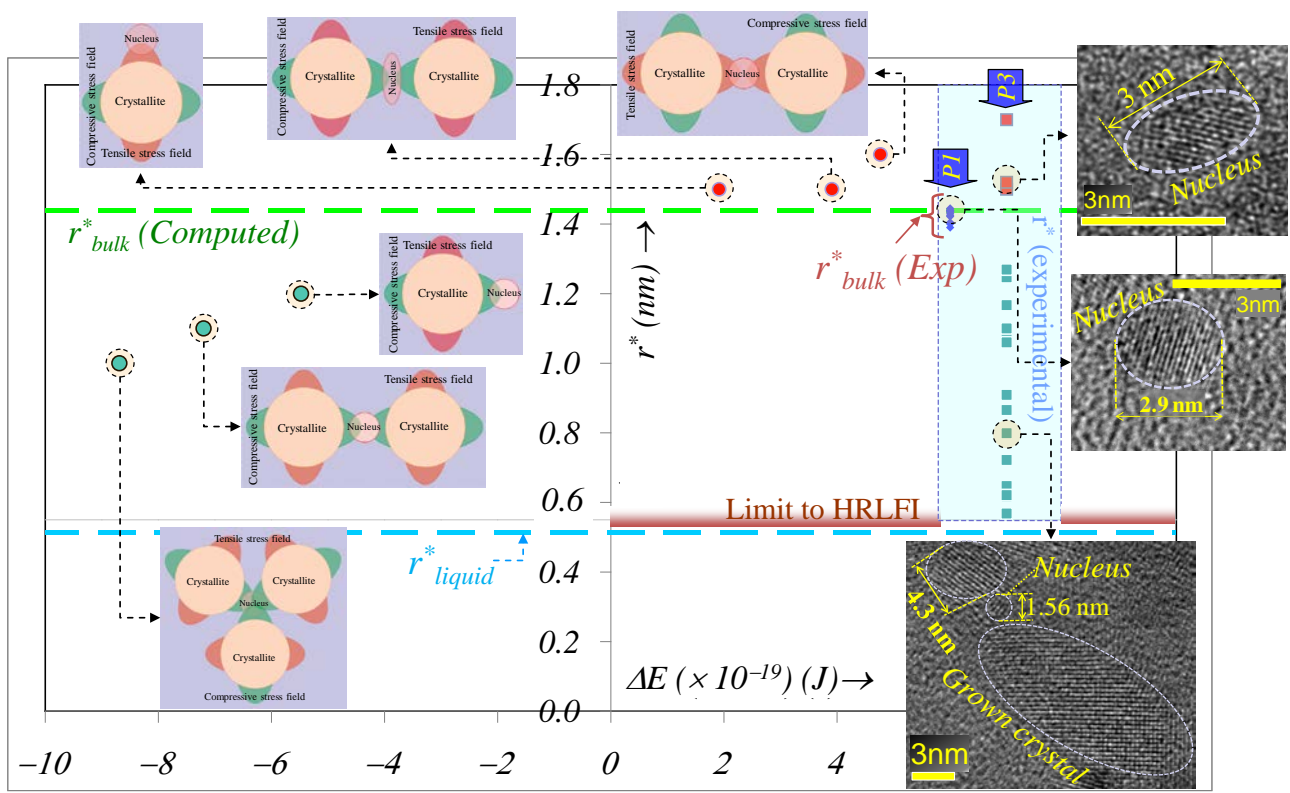

Figure 3. The value of the critical nucleus size ($r^*$) determined using: (i) specific configurations in the computational model (filled circles) and (ii) HRLFI (square or diamond legend in a shaded box). For the computational plot the ordinate is the strain energy difference between nucleation in a strained versus that in an unstrained matrix (ΔE). The experimental samples are of two types: one after P1 treatment and one after P3 treatment. The horizontal dashed lines correspond to $r^*_{bulk}$ (computed), $r^*_{bulk}$ (experimental) and $r^*_{liquid}$. The limit of the HRLFI technique in the measurement of crystal sizes is marked as a broad band (note that this is not the limit of HRTEM, but a practical limit of LFI in the measurement of sizes of crystals). The insets include: schematics of the computational configurations and HRLFI from $Cr_{10}Zr_7$ crystallites. The radii for the experimental crystallite sizes are that of an equivalent circle.

# On the Variability of Critical Size for Homogeneous Nucleation in a Solid-State Diffusional Transformation

Pooja Rani[†], R.M. Raghavendra[†][*] and Anandh Subramaniam[*]

*Materials Science and Engineering, Indian Institute of Technology Kanpur, Kanpur-208016, INDIA, Ph.: +91(512)259 7215*

[†] *Authors with equal contribution and affiliation reflects the place of origin of the work*

[*]*Corresponding author: anandh@iitk.ac.in*

The following aspects related to the manuscript on "Liquid-like Homogeneous Nucleation in Solid-State Diffusional Transformations" are covered in the supplementary material: (i) details of the theoretical, computational and experimental methodology (Section-1), (ii) additional results from the computations and experiments (Section-2) and (iii) detailed discussions regarding models and the experiments (Section-3). A gist of salient results are also presented in the main manuscript.

## 1. Experimental, Theoretical and Computational Methods

Certain details related to the methods used in the main manuscript are described in this section. Much of this methodology forms a part of the standard literature and relevant references have been cited in the main manuscript. To maintain a logical flow, some of the points from the main manuscript are restated here.

### 1.1  Experimental details

The experimental details included pertain to the following: (i) synthesis of samples, (ii) heat treatment, (iii) X-ray diffraction (XRD), (iv) differential scanning calorimetry (DSC) and (v) transmission electron microscopy (TEM).

An alloy of composition $(Cu_{64}Zr_{36})_{96}Al_4$ was prepared from pure elements (Cu, Al & Zr (Crystal bar) of 99.9 wt.% purity) by arc melting in an argon atmosphere. Re-melting of the alloy was carried out four times to ensure homogeneity. The alloy was suction cast in a copper mold to obtain high cooling rate, which can produce a fully amorphous sample. Ti was used as a getter during the melting process to avoid oxidation of the sample. The composition of the amorphous matrix is $(Cu_{64}Zr_{36})_{96}Al_4$ and that of the crystallite is $Cu_{10}Zr_7$. This composition $((Cu_{64}Zr_{36})_{96}Al_4)$ has been reported to have a good glass forming ability, with a critical cooling rate of 40 K/s and critical diameter of 5 mm [1].

X-ray diffraction (XRD) was used to confirm the amorphous structure of the suction cast sample (using CuK$_\alpha$ radiation in a PANalytical Empyrean instrument). The glass transition temperature was determined using differential scanning calorimetry (DSC, TA Instruments, model SDT Q600). DSC was performed under a flow of pure argon at the heating rate 10 K/s.

The following heat treatments are used to obtain partially crystallized samples: (P1) 200°C for 120 min, (P2) 325°C for 30 min and (P3) 200°C for 120 min. The temperatures for the heat treatments are chosen based on the nucleation and growth rate curves (Section-1.2.3) and the DSC results. The sample was sealed in a evacuated quartz tube along with titanium (Ti) getter in order to minimize oxidation during the heat treatment.

In the P1 treatment, the sample is annealed in the nucleation dominated regime to obtain a sparse dispersion of nuclei in an amorphous matrix. The P2 treatment is designed in such a way as to induce the growth of the nuclei which formed during the P1 treatment. This treatment, which is carried out in the growth dominated regime, is used to obtain large/grown crystallites in the size range of 5-15 nm. The P3 treatment, akin to P1, is carried out in the nucleation dominated regime; in order to obtain fresh nuclei in the regions in-between the larger crystallites. The above treatments are based on the following rationale. The nucleation which occurs during the P1 treatment is in a strain free matrix (with little interference between strain fields of the nuclei). On the other hand, some of the nuclei which form during the P3 treatment are in proximity to the grown crystals and hence form in the strain field of pre existing crystallites. A total of seven samples were investigated to obtain HRLFI from nuclei.

A standard technique is used for the preparation of electron transparent samples for transmission electron microscopy [2]. Electropolishing is used for the final stage of the preparation of the thin foil, in order to minimize artifacts related to sample damage (techniques like ion milling are expected to induce some sample damage). The electropolishing solution used was a mixture of 20 vol.% $HNO_3$ and 80 vol.% $CH_3OH$. The twin jet polishing machine was operated at a D.C. potential of 20V and a temperature of the bath was maintained at −30°C.

To calibrate the scale bar of the transmission electron micrographs, a standard gold sample was used. The calibrated scale was used to determine the crystallite size by counting the number of lattice fringes. This was performed by measuring the distance from one bright fringe to another bright fringe. This implies that the technique has an inherent error of about one fringe spacing (i.e. if two dark fringes are present at the end of the 'representative integration line'). Additionally, given that the data is acquired via a digital CCD, an error of at least one digital pixel (~0.032nm) exists. The error bars are not marked in Figure 3 of the main manuscript. The crystallite size is measured using a count of the lattice fringes. This method has an inherent limit of about five bright fringes [3]. If the fringe count is less than this number, then it is difficult to be certain that it is not a 'statistical artifact' (i.e. contrast features occurring by chance which appear to be lattice fringes). Additional aspects are discussed in Section-3.

In the computation of strain energy using the finite element method, the linear misfit is used as an input parameter. The misfit can be calculated using the volume or density mismatch between the glass and the crystal. The density of the glass has been measured using the standard Archimedes method [4]. It is to be noted that the density of the $Cu_{10}Zr_7$ crystal cannot be determined by the same method as that for glass, due to the fact that the crystal transforms to the $Cu_2Zr$ crystal on growth. Hence, the density of the $Cu_{10}Zr_7$ crystal is calculated using the volume of the unit cell and atomic masses ($\rho_{crystal}$ = 7.64 g/cm$^3$).

## 1.2  Theoretical Methods

In this section theoretical calculations are made to calculate: (i) the interfacial free-energy ($\gamma$), (ii) the Gibbs free energy for the transformation ($\Delta G_V$) and (iii) the nucleation and growth rates. The negentropic model has been used for computing the interfacial free-energy [5-7]. Two methods have been used for the calculation of $\Delta G_V$: (a) the Turnbull's method [6] and (b) a first principles based approximation [8]. The classical nucleation theory (CNT) [9] has been used for obtaining the nucleation and growth rates as a function of the temperature.

### 1.2.1  Interfacial Free energy ($\gamma$)

The crystal-glass interfacial free energy ($\gamma$) is a difficult quantity to compute and often it is observed that the value of crystal-liquid interfacial energy is a good enough approximation and can be computed using [5,6]:

$$\gamma = \frac{\alpha \Delta H_f}{(N_A V_m^2)^{1/3}} \qquad (1)$$

where, α is the Turnbull coefficient which depends on the crystal structure, $\Delta H_f$ the enthalpy of fusion, $N_A$ the Avogadro's number and $V_m$ the molar volume of the crystal. The value of $\gamma$ computed using equation (5) is 0.03 J/m$^2$ ($\gamma = 0.03$ J/m$^2$).

*1.2.2 Gibbs Free-Energy ($\Delta G_V$)*

The value of $\Delta G_V$ can be determined using the prescription of Zhou and Napolitano [8]: $\Delta G_V = -16133 - 1.905T$. Another approach for the computation of $\Delta G_V$ is that due to Turnbull [10]: $\Delta G_V = \Delta H_f (T_m - T)/T_m$, where $T_m$ is the melting temperature. It is noteworthy that in the current case both these approaches give a very similar value of $\Delta G_V$.

*1.2.3 Nucleation and Growth Rates ($I_v$ & $U_v$)*

The choice of the temperatures for the heat-treatments P1, P2 and P3, relies on the knowledge of the nucleation and growth rates as a function of the temperature. These rates can be computed using the classical nucleation theory [11]. The nucleation ($I_v$) and growth rate ($U_v$) are given by the following equations [12].

$$I_v = \frac{A_v}{\eta(T)} \exp\left[\frac{-16\pi\sigma^3}{3 K_B T \{\Delta G(T)^2\}}\right] \tag{2}$$

where, $A_v$ is a constant, $\eta$ is viscosity, $\Delta G$ is the difference in the Gibbs free energy difference between the glass and the crystal (per unit volume), $K_B$ is the Boltzmann constant and $\sigma$ is the interfacial energy.

$$U_v = \frac{K_B T}{3\pi l^2 \eta(T)}\left[1 - \exp\left(\frac{-n\Delta G(T)}{K_B T}\right)\right] \tag{3}$$

where, $\eta(T) = A\exp\left(\frac{B}{T-T_0}\right)$, '$l$' is the average atomic diameter and 'n' is the atomic volume ($l = 2.668$ Å and n = 5.549 Å$^3$ for CuZrAl system). The Stoke-Einstein equation has been used in equation (2) to relate diffusivity and viscosity [12,13].

The plot of nucleation and growth rates calculated using equations (1,2) are shown in Figure 1. It is seen from the figure that at 200°C, the rate of growth is negligible and we are in the 'nucleation dominated regime' and at 325°C we are in the growth dominated regime (with a low rate of nucleation). To be on the 'safe side', the treatment intended to cause nucleation is carried out at a temperature lower than that corresponding to the peak nucleation rate temperature and the treatment leading to growth is carried out at a temperature above the peak growth rate temperature.

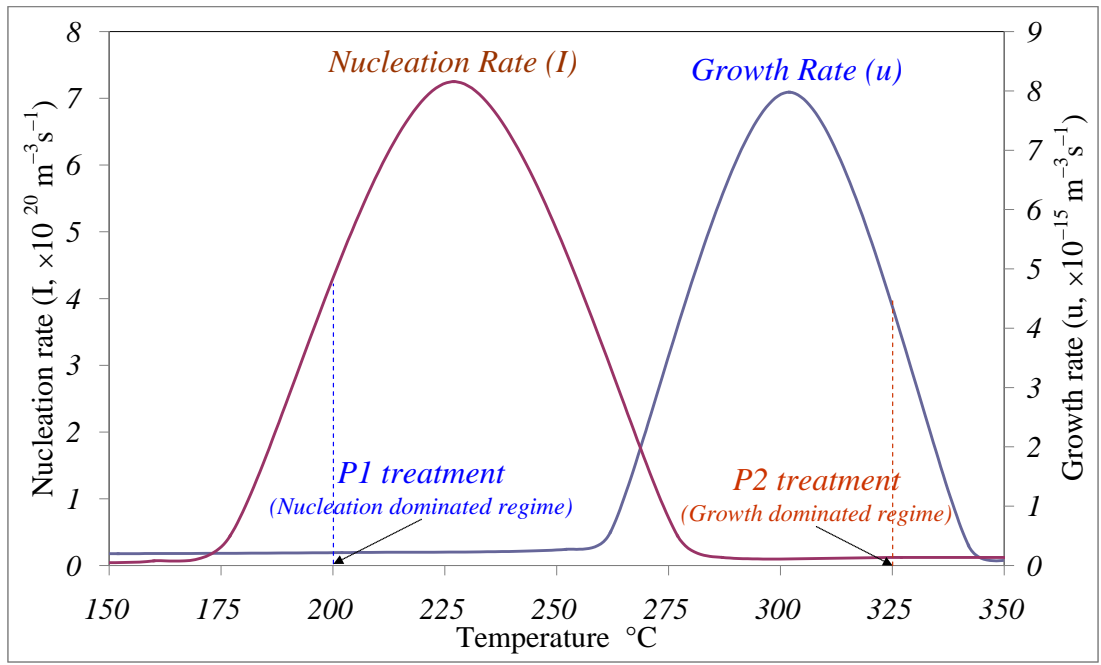

Figure 1. Plots of nucleation and growth rates for the crystallization of the $Cu_{10}Zr_7$ crystal from the $(Cu_{64}Zr_{36})_{96}Al_4$ BMG (plot of equations (1) & (2)). The temperatures for the P1 and P2 treatments are as marked in the figure.

An alternate method to ascertain the relative rates of nucleation and growth at a given temperature, is to utilize the Johnson-Mehl-Avrami-Kolmogrov (JMAK) theory [11] and compute the Avrami exponent using the formula [14]:

$$\left(\frac{dx}{dt}\right)_P = 0.37 n \frac{QE}{RT_P^2} \qquad (4)$$

where, Q, $T_p$, E, R and $(dx/dT)_P$ are the heating rate (K/min), the peak temperature (K), the activation energy (J), the ideal gas constant (J/mole/K) and the maximum crystallization rate respectively. At the annealing temperature of 200°C the value of the exponent (n) is 4.7, thus indicating that we are in the nucleation dominated regime and hence we can be reasonably certain that the crystallites observed in the TEM are 'critical nuclei' (i.e. have not undergone any significant growth).

It should be noted that, the JMAK model used is strictly applicable only in the macroscale, wherein 'random homogeneous nucleation and isotropic growth' conditions exist. Modifications to the model for thin films have been presented in literature, [15-17]; wherein the dimensionality of the transformation has been emphasized. The experimental evidence presented in the current work strongly supports the validity of the JMAK model.

## 1.3 Finite Element Models

The properties of the glass and $Cu_{10}Zr_7$ crystal, which are required as inputs to the finite element models, is as follows. $Y_{glass}$ = 96.4 GPa, $\nu_{glass}$ = 0.355, $Y_{crystal}$ = 131.0 GPa & $\nu_{crystal}$ = 0.326, $\rho_{glass}$ (density) = 7.17 g/cm$^3$ and $\rho_{crystal}$ = 7.64 g/cm$^3$ (volumetric misfit = 6.20%, linear misfit = 2.07%). The elastic constants of the $Cu_{10}Zr_7$ crystal are: $C_{11}$ = 190 GPa, $C_{12}$ = 88 GPa, $C_{13}$ = 102 GPa, $C_{22}$ = 185 GPa, $C_{23}$ = 105 GPa, $C_{33}$ = 167 GPa, $C_{44}$ = 63 GPa, $C_{55}$ = 63 GPa, $C_{66}$ = 47 GPa [18].

The elastic moduli for glass ($Y$ & $\nu$) have been determined by applying the 'rule of mixtures' and is calculated as: $M^{-1} = \sum f_i M_i^{-1}$; where, $M$ & $M_i$ are the elastic constants for glass and the constituent elements respectively and $f_i$ denotes the atomic percentage of the constituent elements [19]. In the context of the elastic properties of the crystal, two types of finite element models are considered: (i) models with anisotropic material properties as inputs and (ii) models where isotropic materials properties. The later models serve to not only highlight issues relevant to the mechanics of the problem, but also serve as reference 'calibration' models. The isotropic material properties of the $Cu_{10}Zr_7$ crystal is obtained from the $C_{ij}$ values using the Voigt averaging method [20]. The material properties used in the current work correspond to that for the bulk material. The determination and utilization of size dependent moduli forms scope for future work.

In the 2D axi-symmetry models the domain is meshed with linear CAX3 elements. In the 3D models the geometry is meshed with hexahedra (C3D8R) and tetrahedra (C3D10) elements. The linear mismatch ($f_m$) is calculated from the volumetric mismatch, which is computed using the density values of the crystal and glass phases. This misfit strain is introduced as thermal strain in the region of the domain corresponding to the crystal(s). The boundary conditions imposed in the models are shown in the respective figures (main manuscript and Section 1.3.1). The size of the domain is kept large enough in comparison to the size of the crystallites, such that it behaves practically as infinite. Mesh convergence is ensured in all cases. The numerical model implemented using Abaqus 2019 software [21].

*1.3.1 Specific Models*

In the main manuscript two finite element models for the homogeneous nucleation of a $Cu_{10}Zr_7$ crystallite in an amorphous (glass) matrix were considered: (i) a single nucleus in an 'infinite' matrix and (ii) a nucleus in the presence of a larger crystal (radius = 5 nm). These two models are expected to capture the 'essential physics' related to the effect of nucleation in a strained matrix, with respect that in a strain-free matrix. In a 'real microstructure', the effects observed for the case of a nucleation in the presence of a large crystal, is expected to be either accentuated or attenuated. I.e., if the nucleus forms in a region of tensile stress, the strain energy penalty will increase; while the converse will be true for nucleation in a compressive region. The additional models considered in this section are to: (i) evaluate the effect of selected parameters or (ii) study the effect of the presence of multiple crystallites; on the magnitude of $r^*$ for nucleation. The models considered are as listed below (Models M1-M7). In some of the cases the same model is utilized with a change in the value of the pertinent parameters.

The relative crystallographic orientation between the 'large crystal' (L) and the nucleus (N) is designated by specifying two parallel basis vectors of the orthogonal set ($[U_1V_1W_1]_L \| [u_1v_1w_1]_N$ & $[U_2V_2W_2]_L \| [u_2v_2w_2]_N$).

*(M1) Size of the pre-existing crystallite (Figure 1a in the main manuscript reproduced as Figure 2).*

In this model the nucleation of a spherical crystal in the proximity of a grown/large crystal is considered. In the main manuscript, the size of the larger crystal was kept constant at 5 nm. In the current model, a larger grown crystal is considered (R = 10 nm, & $r_1 = r_2$). This model is the same that in Figure 1a in the main manuscript.

*(M2) Elastic Isotropic conditions (Figure 3).*

The matrix has an amorphous structure and hence has isotropic properties. In this model the effect of isotropic material properties for the crystal on the value of $r^*$ is evaluated. A 2D axi-symmetric finite element model is used for the computations (Figure 3). This model additionally serves to highlight the key differences between isotropic and anisotropic properties for the crystal on the 'direction' of change of $r^*$ (i.e., an increase or a decrease). In principle model M1 could have been used with isotropic material properties for the crystal. However, we have constructed a separate model so as to serve the purpose of an 'internal check' between 2D and 3D models.

*(M3) The shape of the nucleus (Figure 2).*

The shape of the nucleus is expected to affect the value of $r^*$. In order to keep the analysis simple and tractable, only two additional shapes are considered: the oblate and the prolate ellipsoids. Two values of the ratio of $r_1$ to $r_2$ are considered to obtain a prolate and an oblate spheroidal shape. The values of $r_1/r_2$ chosen are: 0.5 and 2. The relative crystallographic orientations of the nucleus and the large crystal are as shown in the figure.

*(M4) Hydrostatic stress (Figure 4).*

The actual state of stress experienced by a nucleus in the presence of a 'microstructure of large crystals' is expected to be complex. Given that the crystal has a higher density as compared to that of the corresponding glass, the hydrostatic component of the stress imposed by the large crystal plays the key role in determining the strain energy of the system, on the formation of a nucleus in its vicinity. Keeping this in view, the model shown in Figure 4 is developed, wherein radial displacements are used to create conditions of hydrostatic stress of tensile and compressive nature. A 2D axi-symmetric finite element model is used for the computations. This model serves as a 'calibration' of the effect of stress on $r^*$ ($= r^*_{hydrostatic}$).

*(M5) Magnitude of volumetric misfit strain* (Figure 2).

A point of interest is the value of the misfit in determining the magnitude of $r^*$. In this model a hypothetical system is considered, wherein the properties of the matrix and crystallites have been kept constant and the value of the volumetric misfit has been varied. The effect of misfit strain on the magnitude of $r^*$ is computed using the model in Figure 2 (M1), using ansiotropic material properties. The magnitudes of the misfit strains are varied from 2.57% to 0.57% in steps of 0.5%. This model, akin to the M4, is also expected to serve as a model for 'calibration' and reference.

*(M6) Presence of two large crystallites (Figure 5).*

The presence of two large/grown crystals in identical orientation and in proximity is expected to accentuate the effect observed for a single large crystal. The crystallographic orientation of the large crystals is identical and is as marked in the figure. This orientation leads to an overlap of the compressive stress fields of the large crystals. The nucleus has been placed on the inter-nuclear axis of the large crystallites and has a different orientation. The orientation relation between the large crystal and the nucleus is given by: $[\bar{1}\bar{1}2]_L \| [010]_N$, $[\bar{1}\bar{1}0]_L \| [100]_N$. The orientation of the nucleus is such that, the tensile region surrounding the nucleus is aligned in a direction parallel to the inter-crystal axis. The relative orientations have chosen so as to accentuate the effect which was observed (on the magnitude of $r^*$) in the presence of a single large crystallite.

*(M7) Presence of three crystallites (Figure 6).*

This model, with three large crystals in proximity, is designed to further enhance the effect of strain on the value of $r^*$. For simplicity, the three crystallites in model M7 are placed at the vertices of a equilateral triangle. The orientation of one large crystal and the nucleus is as marked in the figure. The other two large crystal are oriented such that the $[\bar{1}\bar{1}2]_L$ is parallel to the line joining the centre of the large crystal to the nucleus. The $[110]_L$ directions of the large crystals are perpendicular to the plane of the drawing.

The results obtained from these models are presented in Section-2.2 and the gist of salient results is presented in main manuscript.

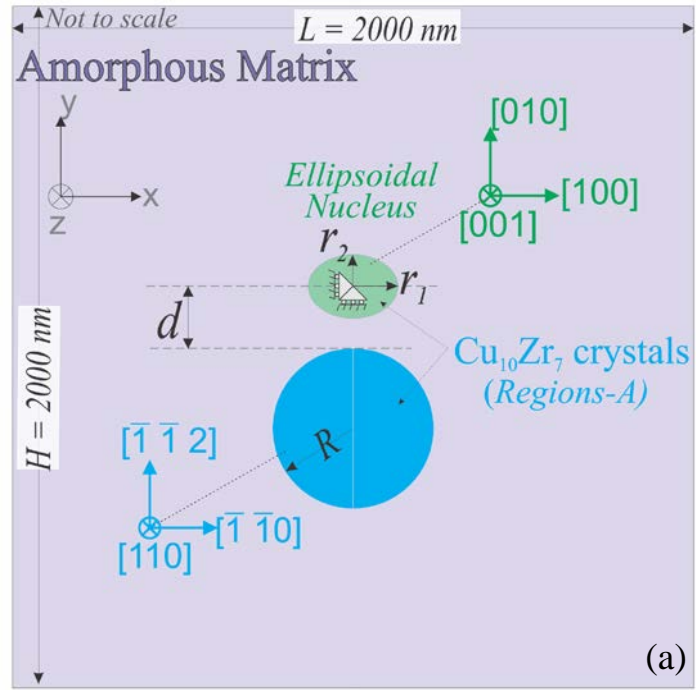

Figure 2. (Model M1). A schematic of the finite element model (2D section of the 3D model) used to simulate the stress state of an spherical nucleus (with $r_1 = r_2$) in the proximity of a pre-existing crystal (of radius R). Eigenstrains are imposed in the Regions-A, corresponding to the volume mismatch between the $Cu_{10}Zr_7$ crystal and the amorphous matrix. The radius of the large crystal (R) is 10 nm. Note that the size of the crystallites has been exaggerated (for better visibility) and hence the figure is not to scale. The nucleus in (ii) is positioned at the centre of the finite element model, but has been shown off-centre in the figure. The central node is locked in x, y & z directions in the finite element model.

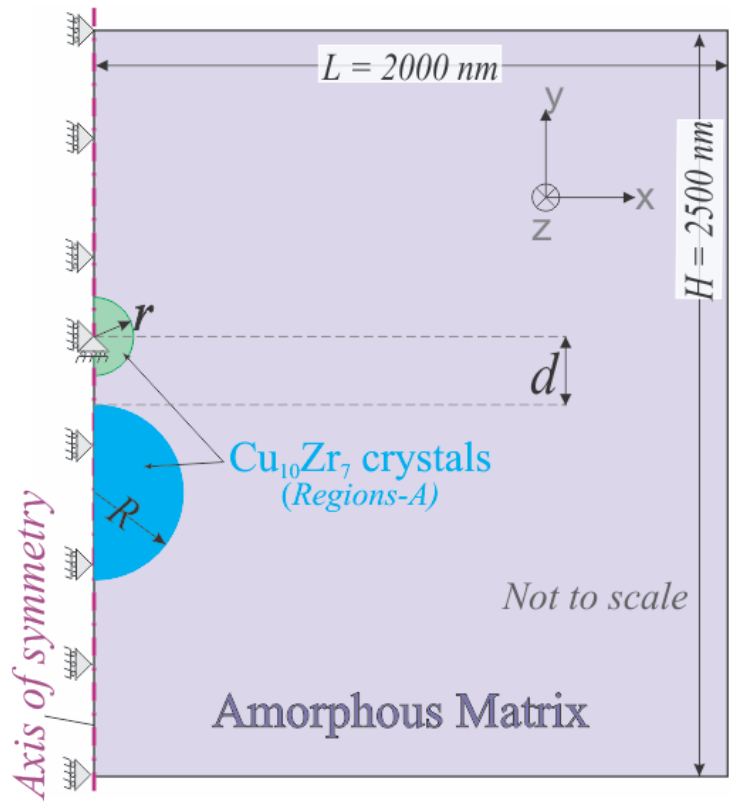

Figure 3. (Model M2). A schematic of the 2D axis-symmetric model used for the study of the effect of isotropic material properties on: (i) the state of stress of a single crystal in an 'infinite' matrix and (ii) the critical size nucleation ($r^*$) in the proximity of a large crystal. Eigenstrains are imposed in Region(s)-A corresponding to the volumetric misfit between the crystal and the amorphous matrix. In the computation of the critical size (part (ii)) the size of the large crystallite is kept constant at 5 nm (R = 5 nm). The size of the nucleus is changed by varying '$r$'. In (ii) the strain energy is computed from the model for different values of '$d$'. The boundary conditions are as marked in the figure.

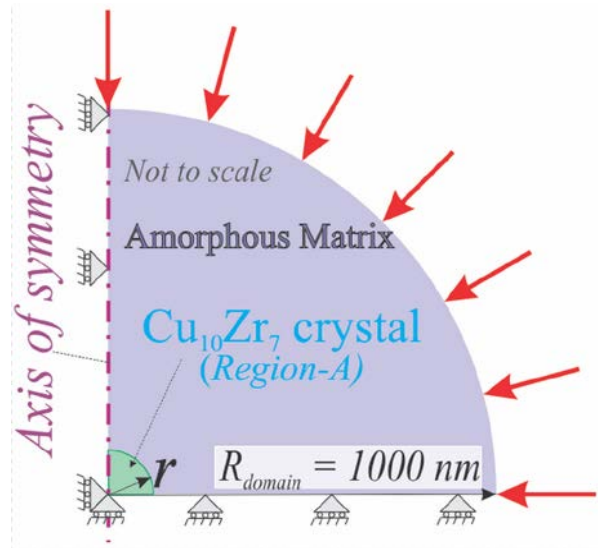

Figure 4. (Model M4). A schematic of the 2D axis-symmetric model used for the study of the effect of hydrostatic stress on the critical size of the nucleus. Displacements are imposed along the radial directions in order to induce hydrostatic stresses in the body. The direction of the imposed displacements, in order to obtain a state of hydrostatic compression in the crystal, is shown as red vectors. The directions of the displacements are reversed to obtain a state of tensile stress in the crystal. The 'mirror' boundary conditions along the abscissa and the axi-symmetry along the ordinate are to be noted.

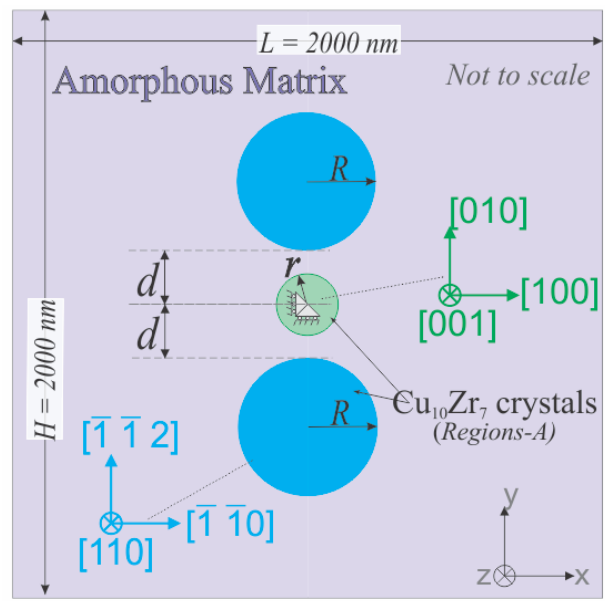

Figure 5. (Model M6). A 2D schematic (mid-section along the z-axis) of the 3D finite element model used to evaluate the effect of the presence of two large crystals in proximity on the $r^*$ for nucleation. The orientations of the large crystallites are identical. The radius of the large crystals is 5 nm (R = 5 nm).

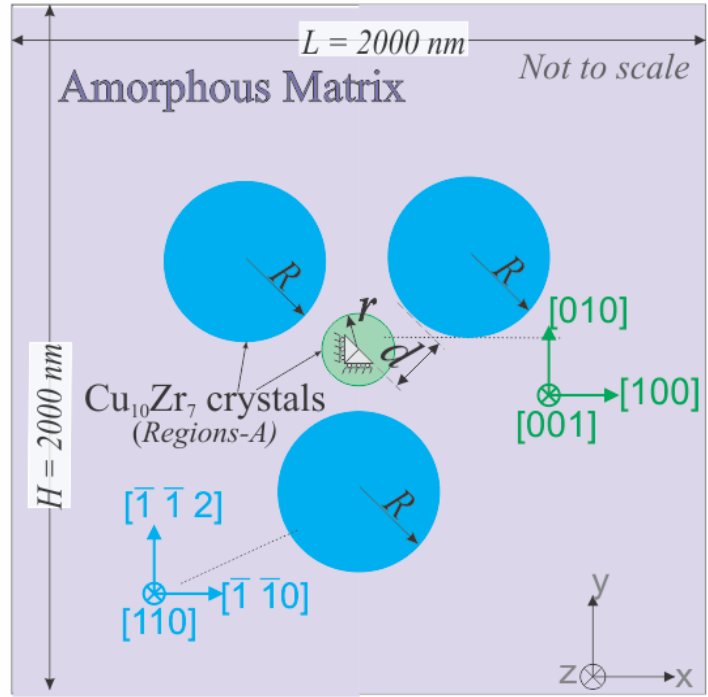

Figure 6. (Model M7). A 2D schematic (mid-section along the z-axis) of the 3D finite element model used to evaluate the effect of three large crystals on the $r^*$ for nucleation. The orientation of one large crystal and the nucleus are marked in the figure. The other two large crystals are oriented such that the $[\bar{1}\bar{1}2]_L$ is parallel to the line joining the centre of the large crystal to the nucleus. The radius of the large crystals is kept constant at 5 nm. The central node is locked in x, y & z directions. The three large crystals are located at the vertices of equilateral triangle with an edge length of 12 nm. The nucleus is positioned at the centroid of the triangle.

## 2. Results

### 2.1 Experimental Investigations

In this section we present the following: (i) results from the suction cast $(Cu_{64}Zr_{36})_{96}Al_4$ alloy (XRD pattern, TEM results, DSC results), (ii) HRLFI from the partially crystallized alloy.

Figure 7 shows the XRD pattern obtained from the suction cast $(Cu_{64}Zr_{36})_{96}Al_4$ alloy. The broad peak indicates the formation of an amorphous structure. The non-existence of crystalline phases is confirmed by the selected area diffraction (SAD) pattern and high resolution transmission electron microscopy (Figure 8). The absence of lattice fringes (Figure 8b) and the presence of only diffuse rings in the SAD pattern (Figure 8a) confirms the formation of a fully amorphous sample. This is to be contrasted with the SAD pattern

*12*

obtained from the annealed sample (200°C for 120 min) as shown in (Figure 8c). A study of the figure establishes that, the glass (as-cast sample) does not have any quenched-in nuclei (pre-existing crystallites) and that, the nuclei (crystallites) formed only on annealing of the sample.

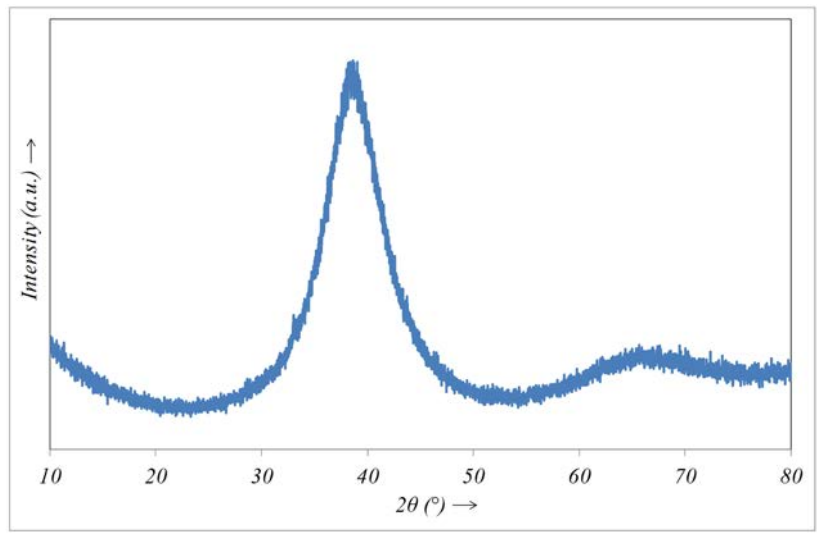

Figure 7. XRD pattern with a broad peak showing the formation of an amorphous structure in the suction cast $(Cu_{64}Zr_{36})_{96}Al_4$ alloy.

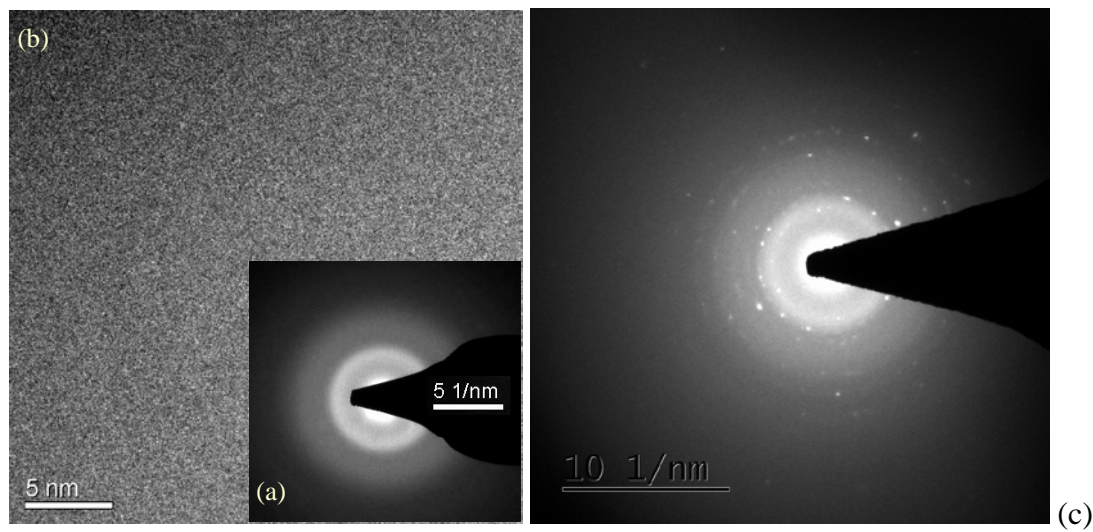

Figure 8. (a) (Inset) SAD pattern obtained from the suction cast $(Cu_{64}Zr_{36})_{96}Al_4$ alloy showing diffuse rings. This confirms the existence of only the amorphous phase. (b) High resolution transmission electron microscopy micrograph showing 'salt & pepper' contrast, which is typical of the amorphous phase. The absence of lattice fringes is to be noted. (c) SAD pattern obtained from the annealed sample (200°C for 120 min) indicating the presence of crystallites.

Figure 9 shows the DSC plot obtained at a heating rate of 10 K/s, wherein the glass

*13*

transition ($T_g$) and the crystallization ($T_x$) temperatures are marked. Annealing of the samples were performed below the glass transition temperature.

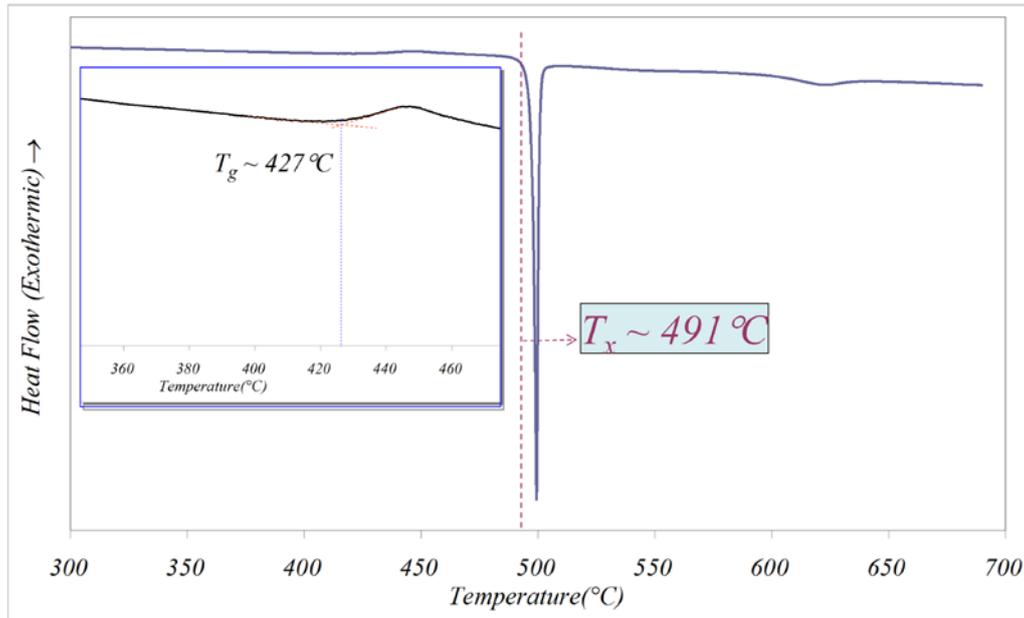

Figure 9. DSC plot used for the determination of the glass transition temperature ($T_g$). $T_x$ is the temperature of the onset of crystallization.

The density of glass is determined by Archimedes method is ($\rho_{glass}$ =) 7.17 g/cm$^3$. The density of the $Cu_{10}Zr_7$ crystal is ($\rho_{crystal}$) = 7.64 g/cm$^3$.

## 2.2    Computational Investigations

In this sub-section the results related to the finite element simulations are presented. These results are generated using the models considered earlier.

The value of $r^*$ for R = 5 nm and d = 2 nm was computed to be 1.2 nm. On increasing the size of the large crystal to 10 nm (= R), keeping d = 2 nm, the value of $r^*$ decreased to 1.1 nm. The hydrostatic stress contours corresponding to these two cases is shown in Figure 10. It is seen that a larger crystallite gives rise to a larger volume of compressive stress in matrix. This also applies to a certain bandwidth of compressive stress of high magnitude (e.g. the region occupied by compressive stress of magnitude in the range −0.1 to −0.53 GPa is larger for the bigger crystal). Hence, the nucleus can be accommodated in the high magnitude compressive region of the larger crystal with relative ease as compared to the smaller one. This implies that with a progress of the transformation, as the size of the crystallites increases, the magnitude of $r^*$ can display a higher magnitude of decrease (for nucleation in the 'favourable' region).

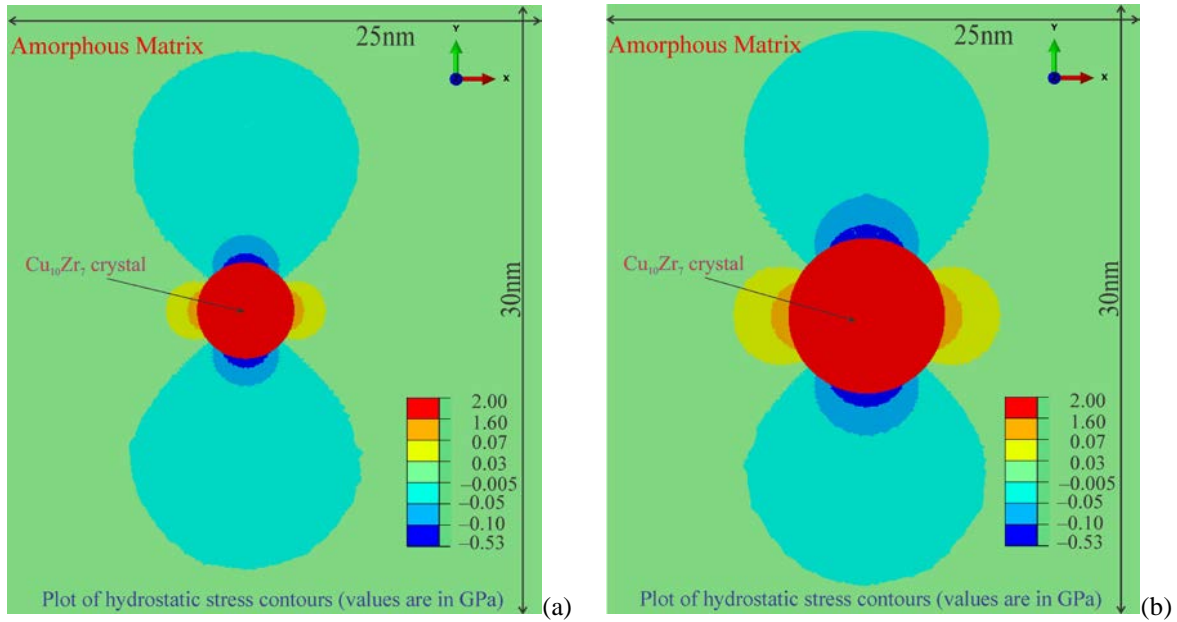

Figure 10. Plot of hydrostatic stress contours for two sizes of large crystals (R =5 nm & R = 10 nm). Note that the magnitude of the highest compressive stress, along with the stressed region, shows an increase.

The plot of hydrostatic stress for the case of a nucleus in the presence of larger crystallite (R = 5 nm) was presented as an inset to Figure-1c in the main manuscript. Figure 11 shows the plot of $\sigma_{xx}$ and $\sigma_{yy}$. The crystallites are in a state of tensile stress (in both $\sigma_{xx}$ and $\sigma_{yy}$ plots). As expected, the approximate religions of the tensile and compressive lobes are reversed between the two figures (Figure 11a versus Figure 11b). It is to be noted that, in the proximity of the nucleus the sign of stress may be reversed (i.e. compressive regions may be rendered tensile) or may be accentuated (i.e. compressive regions may become more compressive).

*15*

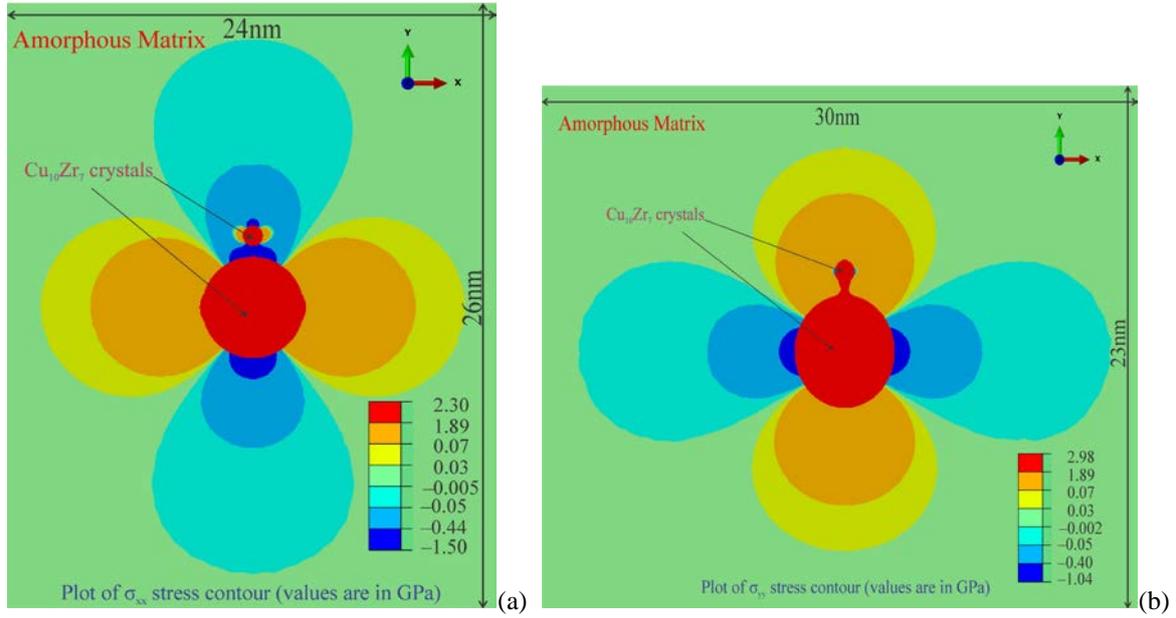

Figure 11 A plot of $\sigma_{xx}$ (a) and $\sigma_{yy}$ (b) stress contours computed using the model shown in Figure 1a of the main manuscript. It is to be noted that the sign of $\sigma_{xx}$ is reversed in the proximity of the nucleus (i.e. the stress is tensile inside the nucleus and in certain proximal regions of the nucleus).

The results from Model-(M2) are shown in Figure 12. Figure 12a shows a plot of hydrostatic stress contours in the presence of a crystal of size 5 nm and a nucleus of size r = 1.6 nm at a distance d = 2 nm. A plot of variation of strain energy ($E_{strain}$) and $r^*$ with *'d'* is shown in Figure 12b. The following points are to be noted from the figure. (i) If isotropic elastic properties are used for the crystal, the regions around the same are in a state of tensile stress (i.e. regions of hydrostatic compression are absent). (ii) The formation of the nucleus always leads to an increase in the strain energy of the system. (iii) The energy cost for the formation of the nucleus increases monotonically with a decreasing 'd'; with an increasing slope. (iv) The value of $r^*$ increases with decreasing *'d'* and for a value of d < 2 nm $r^*$ diverges.

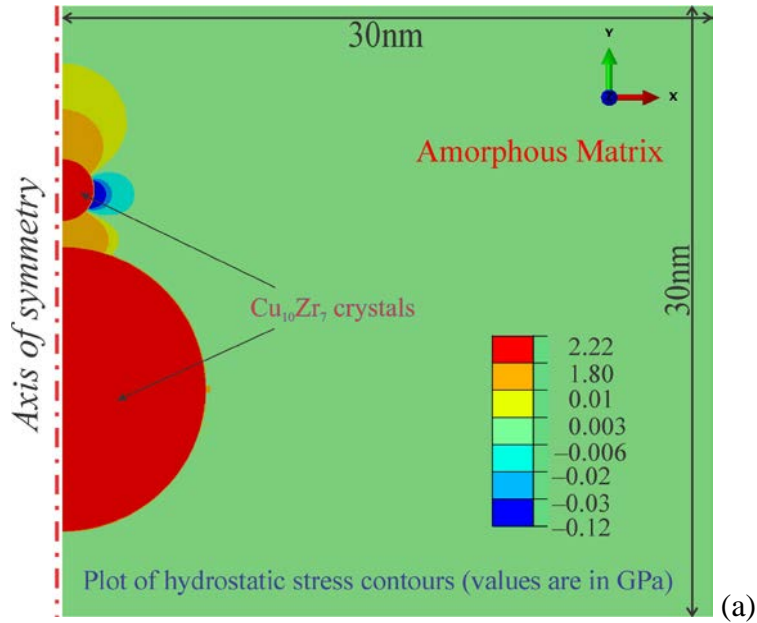

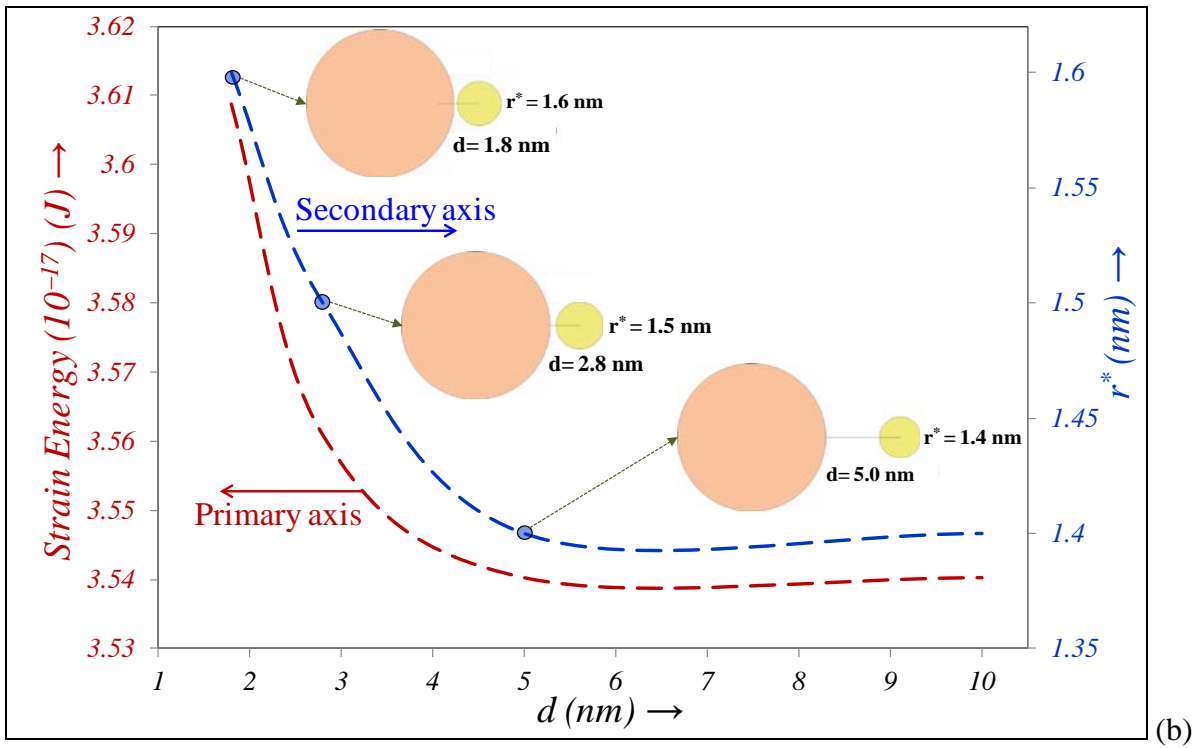

Figure 12. Results obtained from Model-M2. (a) Plot of hydrostatic Stress contours in the presence of a crystal of size 5 nm and a nucleus of size r = 1.6 nm at a distance d = 2 nm. The absence of compressive regions around the large nucleus is to be noted. (b) Plot of variation of strain energy ($E_{strain}$) and $r^*$ with 'd'. It is to be noted that $r^*$ steeply increases with decreasing 'd'. Insets schematically represent the relative positions of the crystals at three values of 'd'.

Figure 13 shows the plot of hydrostatic stress in the presence of a crystal of R = 5 nm and a nucleus, which is either a oblate or prolate ellipsoid/spheroid (Model-M3). The nucleus is positioned in the compressive region of the large crystal at a distance (*d*) of 4 nm. In order to make a comparison with the spherical nuclei, an equivalent radius ($r_{equivalent}$ ($r_{eq}$)) is computed for these shapes using the formula: $\frac{4}{3}\pi r_{eq}^3 = V_{spheroid}$. Thus, the magnitude of $r^*$ computed corresponds to this equivalent radius ($r_{eq}^* = r^*$). The value of $r^*$ for an oblate spheroid (with $r_1/r_2 = 1/2$) and a prolate spheroid (with $r_1/r_2 = 2/1$) are: 1.4 nm and 1.3 nm, respectively. The corresponding value of a spherical nucleus is 1.5 nm (d = 4 nm). Note that a large value of '*d*' has been chosen to accommodate both the shapes of the nuclei. The depression in the magnitude of the $r^*$, for the cases with oblate and prolate ellipsoid nuclei, with respect to a spherical nucleus, is to be noted. This decrease is more for the prolate spheroid, wherein a larger region of the nucleus is positioned within a compressive region of the large crystal (with a higher magnitude of compressive stress). This implies that the shape of the nucleus plays an important role in determining the magnitude of $r^*$. This further implies that the details of the overlap of the stress fields of the nucleus with that of the matrix, which arises from microstructural details existing at any point of time during the phase transformation, is a key factor in determining the magnitude of $r^*$.

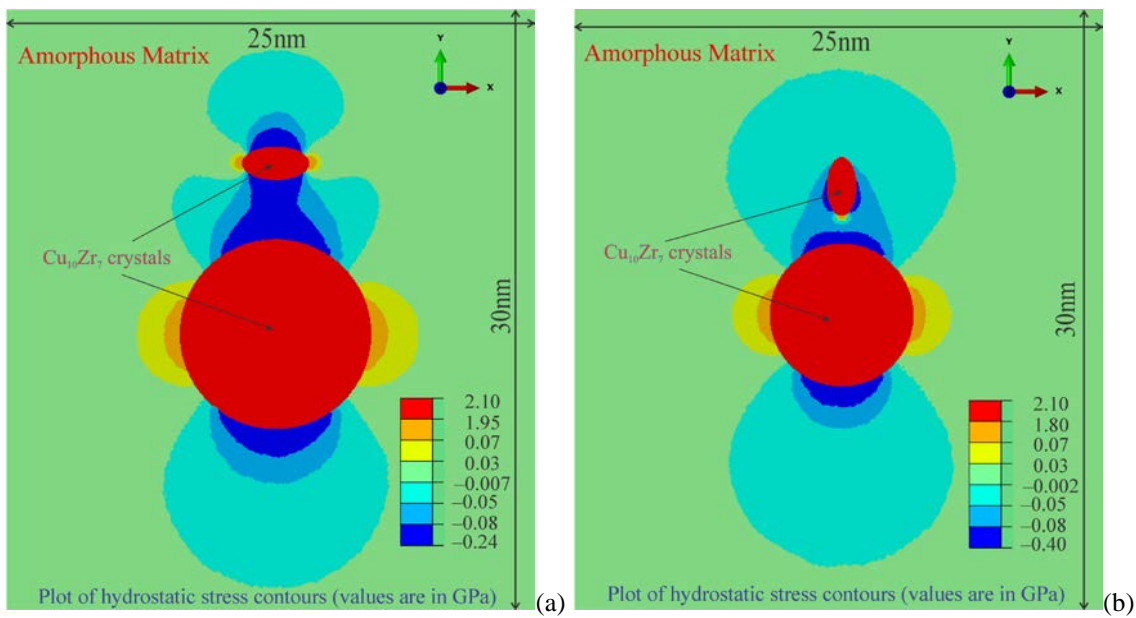

Figure 13. The results are obtained using Model-(M3). Plot of hydrostatic stress contours in the presence of a spherical crystal of size R = 5 nm and a ellipsoidal nucleus. In (a) the nucleus is an oblate ellipsoid with $r_1/r_2 = 1/2$. In (b) the nucleus is a prolate ellipsoid with $r_1/r_2 = 2/1$. The distance d = 2 nm in both (a) and (b).

The results obtained from Model-M4 is shown in Figure 14. The variation of critical size

($r^*_{hydrostatic}$) as a function of the hydrostatic stress is shown in the figure. It is to be noted that the hydrostatic stress is generated by the radial displacements imposed in Model-(M4). As expected, the value of $r^*_{hydrostatic}$ decreases in the presence of hydrostatic compression and increases in the presence of hydrostatic tension. The magnitude of $r^*$ for zero stress corresponds to $r^*_{bulk}$. This results unequivocally establishes the role of hydrostatic stress on the strain energy cost for the formation of the nucleus and hence the value of the critical size.

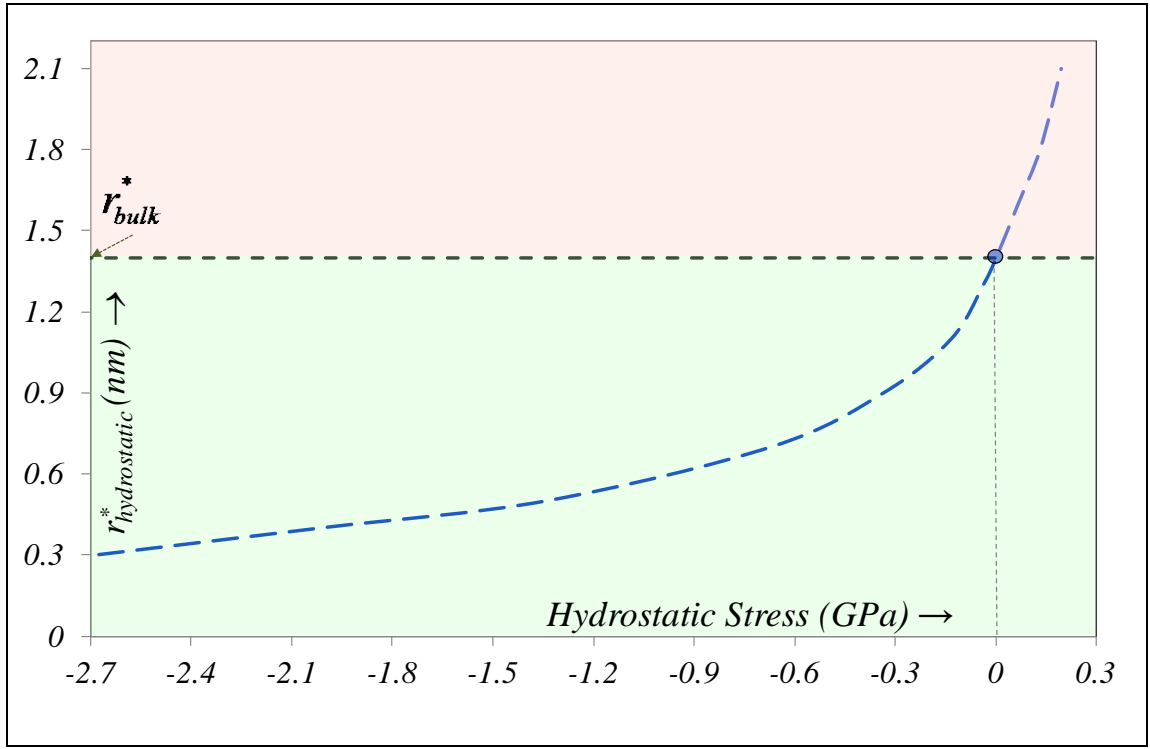

Figure 14. Results corresponding to Model-(M4). Plot of critical size ($r^*_{hydrostatic}$) as a function of the hydrostatic stress. In the absence of any stress the magnitude of $r^*$ corresponds to $r^*_{bulk}$ (= 1.4 nm).

Figure 15 shows the plot of $r^*$ as a function of the misfit ($f_m$) between the crystal and the amorphous matrix. It is to be noted that the magnitude of the misfit is used as a free parameter ('artificially'), to isolate its effects on the value of $r^*$. The model used for the computations is the 3D anisotropic model (Model-M1). With increase in the misfit, the strain energy cost for the formation of a nucleus in an unstrained matrix increases, thus leading to an increase in $r^*$. However, the important factor to be evaluated is the effect of misfit on the relative benefit of nucleation in a strain free matrix versus that in the proximity of a large crystal. Data in this regard is summarized in Table I, for the following parameters: R = 5 nm and d = 2 nm (nucleus is spherical and is positioned in the compressive region of the large crystal). It is seen that the relative change in $r^*$ increase with the value of the misfit. I.e., although the magnitude of $r^*$ increases with misfit, the relative benefit for a suitably positioned nucleus is higher. It is to be noted that the magnitude of $r^*$ approached that for nucleation from a liquid (i.e zero strain energy penalty) for a misfit value of 0.6 %.

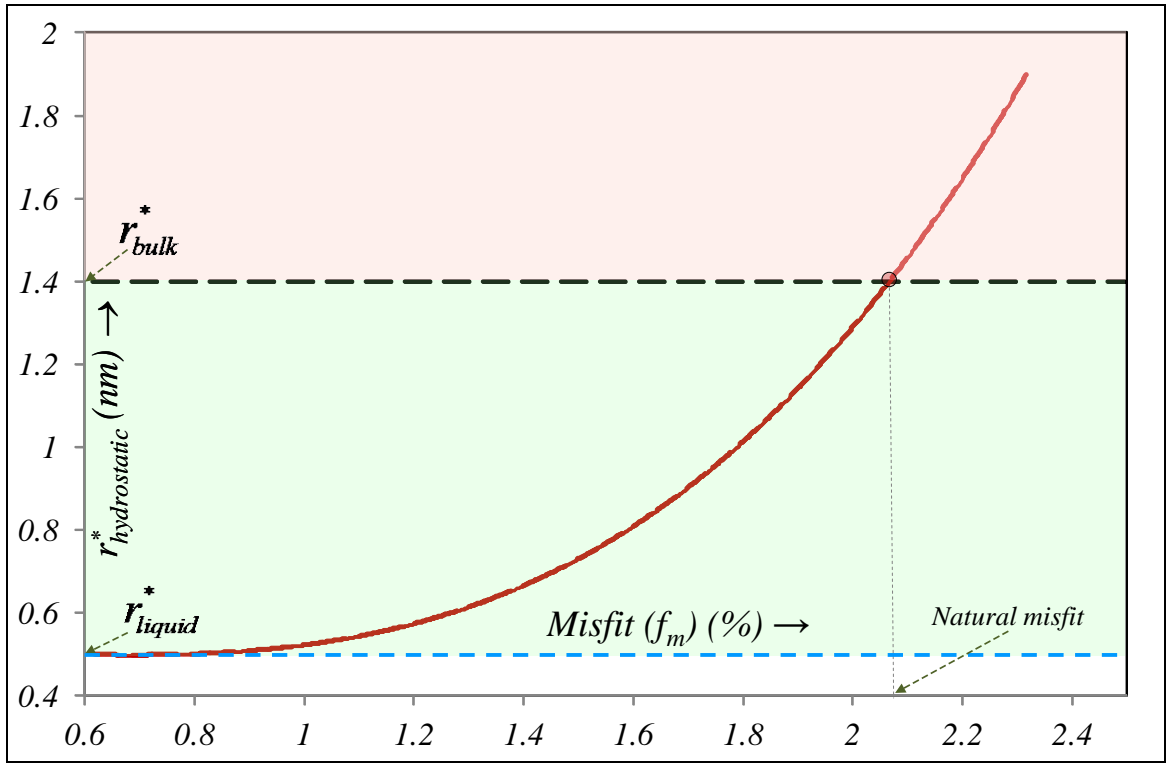

Figure 15. The result is obtained using model-M1. The plots show the variation in $r^*$ as a function of the misfit ($f_m$). The horizontal dashed lines correspond to $r^*_{bulk}$ & $r^*_{liquid}$.

Table I. The effect of the magnitude of the misfit ($f_m$) on the relative benefit of nucleation in a strained matrix with respect to that in a strain free matrix. The model-M1 used in the computations (with R = 5 nm, d = 2 nm). Misfit is the only parameter which is varied, keeping the others constant. A negative value for $\Delta r^*$ (in %) implies an increased benefit due to the presence of a large crystal.

| Misfit ($f_m$), % | $r^*_{bulk}$ [nm] | $r^*$ (model-M1) [nm] | $\Delta r^*$ % | Comments |
|---|---|---|---|---|
| 1.27 | 0.7 | 0.6 | −14.29 | Reduced misfit with respect to the natural value |
| 2.06 | 1.4 | 1.1 | −21.42 | Using 'natural' values of misfit |
| 2.31 | 2.5 | 1.7 | −32.00 | Increased relative 'benefit' |

The results from Model-M6, wherein two large crystals accentuate the effect of the presence of a single large crystal, is shown in Figure 16. The figure shows a plot of hydrostatic stress contours in the presence of two crystals (R = 5 nm) and a nucleus (r = 1.1 nm). The distance *'d'* is 2 nm. The nucleus is symmetrically positioned in the compressive regions of the two large crystals. The value of $r^*$ computed for this configuration is 1.1 nm, which implies a significant depression in the value of $r^*$ as compared to that for nucleation in a strain free matrix ($r^*_{bulk} = 1.4\ nm$).

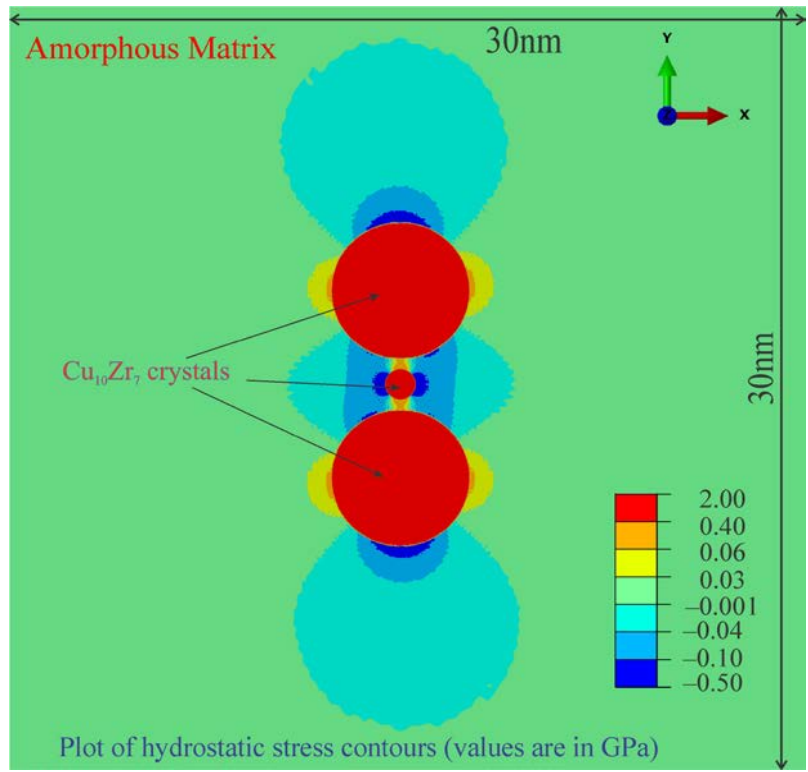

Figure 16. The results correspond to Model-M6. Plot of hydrostatic stress contours in the presence of two crystals (R = 5 nm) and a nucleus (r = 1.1 nm). The distance *'d'* is 2 nm.

Figure 17 shows a plot of hydrostatic stress contours in the presence of three large crystals at the vertices of an equilateral triangle and a nucleus at its centroid (corresponding to Model-M7). The value of $r^*$ computed for this configuration is 1 nm a further decrease over the case of two large crystals. This implies that, in realistic microstructures there can arise regions of accentuated compressive stress arising from multiple crystallites; which may lead to a further decrease in the magnitude of $r^*$ (i.e., these regions are preferred regions for nucleation). This aspect corroborates well with the experimental observations, where a significant depression in the magnitude of $r^*$ was observed.

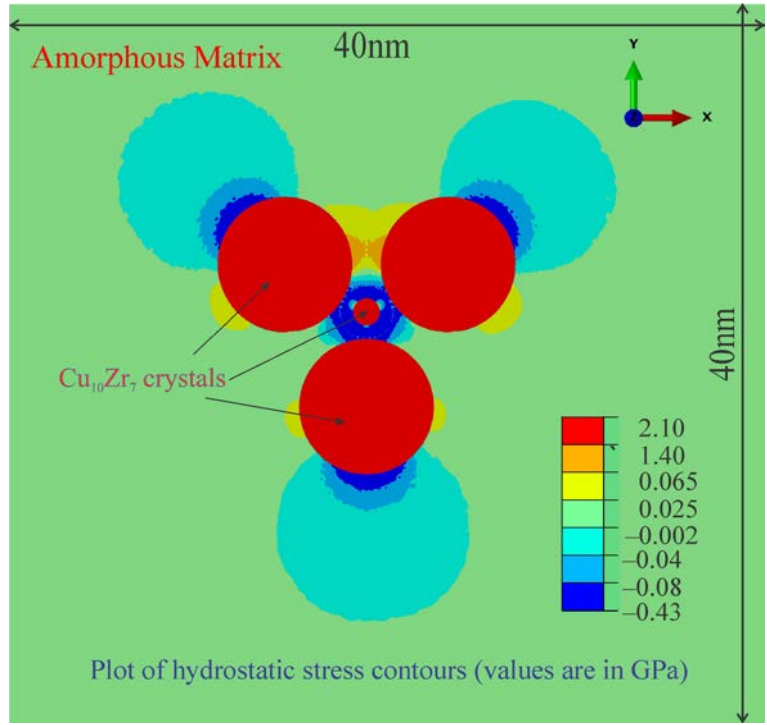

Figure 17. The results correspond to Model-M7. Plot of hydrostatic stress contours in the presence of: (a) three crystals (R = 5 nm) and a spherical nucleus (r = 0.9 nm) at a distance d = 2 nm from each of the crystallite.

## 3. Discussions (& Additional Results)

A discussion of few important points is warranted here. These pertain to: (i) alternate methods for the determination of $r^*$, (ii) the use of an appropriate nucleation theory, (iii) the values of the material properties used in the computations, (iv) the use of HRLFI for the measurement of crystallite sizes and (v) the broader applicability of the conclusions.

Ab-initio density functional theory (DFT) is an sophisticated technique for atomic level computations. However, for the systems sizes involved in the current set of computations, this approach will prove to be 'computationally prohibitive'. On the other hand, molecular dynamics (MD) can handle larger systems sizes; but, the timescales that the technique can handle is typically of the order of microseconds [22]. Researchers have come up with acceleration techniques and enhanced sampling methods to access longer timescales [23,24]. However, given that the timescales involved in 'rare events' like nucleation exceeds milliseconds, this requirement lies well beyond a routine computation. This is especially true for the kind of work in the current investigation, wherein more than one crystallite is involved. The availability of reliable interatomic potentials for alloys (binary and multi-component) is another challenge with the use of MD. The fruitful use of these techniques to accurately determine the magnitude of $r^*$ (for a given microstructure), forms scope for future work.

The accurate determination of the interfacial energy is a difficult task, even for a planar crystal-crystal interface. Hence, the computation of crystallite size and orientation dependent glass-crystal interfacial energy is a really challenging one. This needs to be taken up in the future to obtain quantitatively accurate values of $r^*$. The classical nucleation theory (CNT), which is at the heart of the concept of 'statistical random fluctuations', has served as a good theory to explain nucleation in diverse systems for more than 150 years [9]. The applicability of this theory has been investigated and alternate formulations, including the 'non-classical nucleation theory' have been proposed [25-27]. These theories are motivated by observations in cases, where the nucleation is a multistage process [28]. This is also termed as the Ostward rule of stages and seems to be important in cases, wherein liquid or a gaseous phases is involved. The first step in these scenarios could involve a cluster ('pre-nucleation cluster') [25] or an amorphous phase [29]. The current understanding in this regard is that, the CNT can be successfully applied to many systems; including those which are more complex than the one in the current study, without a significant error [9, 29-31]. In the present work, it is observed that the value $r^*$ computed using CNT matches well with that experimentally measured using HRLFI.

Material properties like elastic constants, interface energy, etc., are expected to be size dependent, especially in the nanoscale regime. In the current work 'macroscale'/bulk properties have been used, which has been necessitated by the lack of sufficiently accurate computations of the size dependent values of these parameters. Improvement on the assumptions used in the models, along with the use of accurate values for the material properties for the computation of $r^*$, forms scope for the future work.

As discussed in the main manuscript, the devitrification of glass serves as a model system to study nucleation behaviour, in a solid state phase transformation. One important reason for this relates to the use of HRLFI for the measurement of crystallite sizes. However, this technique imposes a lower limit to the size of a crystallite which can be measured. This lower limit is set by microscopic and crystallographic variables. A practical limit of about 0.55 nm is shown as broad band in Figure 3 of the main manuscript. For the $Cu_{10}Zr_7$ crystal imaged using the Titan $G^2$ 60-300 HRTEM, the (10 0 2) was the highest order crystal plane which was imaged. This has *'d'* spacing of 1.22 Å. Assuming that a region having five bright fringes can be definitively identified as a crystal, we arrive at the figure of 0.55 nm as a practical limit for identifying crystals in the current work. It is to be noted that, the smallest size of the nuclei which are reported in the main manuscript, are close to this lower limit. This implies that any smaller crystallites would go undetected by this technique. A related 'philosophical' question is: "How large has an ordered region to be, before it can actually be considered as a crystal?". The current manuscript has taken a practical standpoint on this issue.

The key conclusion of the current investigation is that, the critical radius for nucleation ($r^*$) is a time dependent variable for a solid state diffusional transformation. The assumption that $r^*$ is a cardinal constant is disproved by using the example of the formation of the $Cu_{10}Zr_7$ crystal during the devitrification of a Cu-Zr-Al glass. This system was chosen purely as a model system for the reasons listed in the main manuscript, without any expectation or bias. This alludes to the possibility that, the conclusions drawn have a broader applicability to a variety of solid to solid diffusional transformations. Three key parameters determine if the magnitude of $r^*$ will change with the progress of the transformation: (i) the elastic properties of the parent and the product phases, (ii) the misfit between the parent and the product phases and (iii) the magnitude of interfacial energy.

We have already established that isotropic elastic properties leads to an increase in the magnitude of $r^*$ and anisotropy is essential for a decrease. To test the broader applicability of the conclusions, we perform a series of model computations of $r^*$; a summary of which can be found in Table II. In these set of computations, a single parameter is varied, keeping the others constant. The relative benefit of nucleation in a strained matrix is listed as $\Delta r^* = \left( \dfrac{r^* - r^*_{bulk}}{r^*_{bulk}} \right) \times 100$ in the table. The value of $r^*_{bulk}$ is computed using a new set of material properties. A negative value of $\Delta r^*$ arises from a strain energy benefit due to the presence of a large crystal in proximity. The value of $r^*$ in the aforementioned equation is computed for each set of properties appearing in the table. The figure corresponding to the computation is listed in the last column of the table.

Table II. The effect of the magnitude of various parameters on the value of $r^*$. $\Delta r^*$ is the difference between the nucleation in a strain free matrix, with respect to that in a strained matrix (Model- M1/M2, R = 5 nm, d = 2 nm). The data for the effect of misfit for the computation of $\Delta r^*$ is reproduced from Table I. Column-3 corresponds to the natural value of a given parameter and columns-2 & 4 to the altered values.

| $r^*_{bulk} = 1.4\ nm$ | | $r^*_{bulk}$, $\Delta r^*$ | | Model Figure |
|---|---|---|---|---|
| Parameter | (Decreased value) | (Value for the current system) | (Increased value) | |
| Young's Modulus of matrix ($Y_{glass}$) (Isotropic material properties) | ($Y_{glass}$ = 49 GPa) $r^*_{bulk}$ = 0.8 nm $\Delta r^*$ = 12.50% | ($Y_{glass}$ = 98 GPa) $r^*_{bulk}$ = 1.4 nm $\Delta r^*$ = 7.14% | ($Y_{glass}$ = 102.9 GPa) $r^*_{bulk}$ = 1.5 nm $\Delta r^*$ = 6.67% | M2 Figure 3 |
| Young's Modulus of crystal ($Y_{crystal}$) (Isotropic material properties) | ($Y_{crystal}$ = 120 GPa) $r^*_{bulk}$ = 1.3 nm $\Delta r^*$ = 7.69% | ($Y_{crystal}$ = 131 GPa) $r^*_{bulk}$ = 1.4 nm $\Delta r^*$ = 7.14% | ($Y_{crystal}$ = 137.6 GPa) $r^*_{bulk}$ = 1.5 $\Delta r^*$ = 6.67% | M2 Figure 3 |
| Misfit ($f_m$) | ($f_m$ = 1.27%) $r^*_{bulk}$ = 0.7 nm $\Delta r^*$ = −14.29% | ($f_m$ = 2.06%) $r^*_{bulk}$ = 1.4 nm $\Delta r^*$ = −21.42% | ($f_m$ = 2.31%) $r^*_{bulk}$ = 2.5 nm $\Delta r^*$ = −32.00% | M1 Figure 2 |
| Interfacial energy ($\gamma$) | ($\gamma$ = 0.0189 J/m$^2$) $r^*_{bulk}$ = 1 nm $\Delta r^*$ = −30.00% | ($\gamma$ = 0.0269 J/m$^2$) $r^*_{bulk}$ = 1.4 nm $\Delta r^*$ = −21.42% | ($\gamma$ = 0.0404 J/m$^2$) $r^*_{bulk}$ = 2.1 nm $\Delta r^*$ = −14.29% | M1 Figure 2 |

To study the effect of anisotropy of the magnitude of $r^*$, we have chosen the universal anisotropy parameter ($A^U$) [32]. This is a generalized parameter, which is also applicable to non-cubic crystals and is given by [32]: $A^U = 5\left(\dfrac{G^V}{G^R}\right) + \left(\dfrac{K^V}{K^R}\right) - 6$, where $G$ and $K$ are the shear & bulk moduli and the superscripts $V$ & $R$ refer to Voigt and Reuss averaging of the elastic constants. Other parameters have also been developed in literature to capture the anisotropy of crystals [33]. As value of $A^U$ approaches zero, the system tends towards isotropy and a value of six for the parameter corresponds to the highest anisotropy. The value of $A^U$ is varied by changing $C_{11}$. It can be seen from Figure 18 that as the anisotropy is increased, not only that the magnitude of the highest compressive stress increases, but also the volume of the highly stress region. This implies that an increased anisotropy will lead to a larger decrease in the magnitude of $r^*$ for a nucleus at a constant distance from the large crystal (within the compressive field).

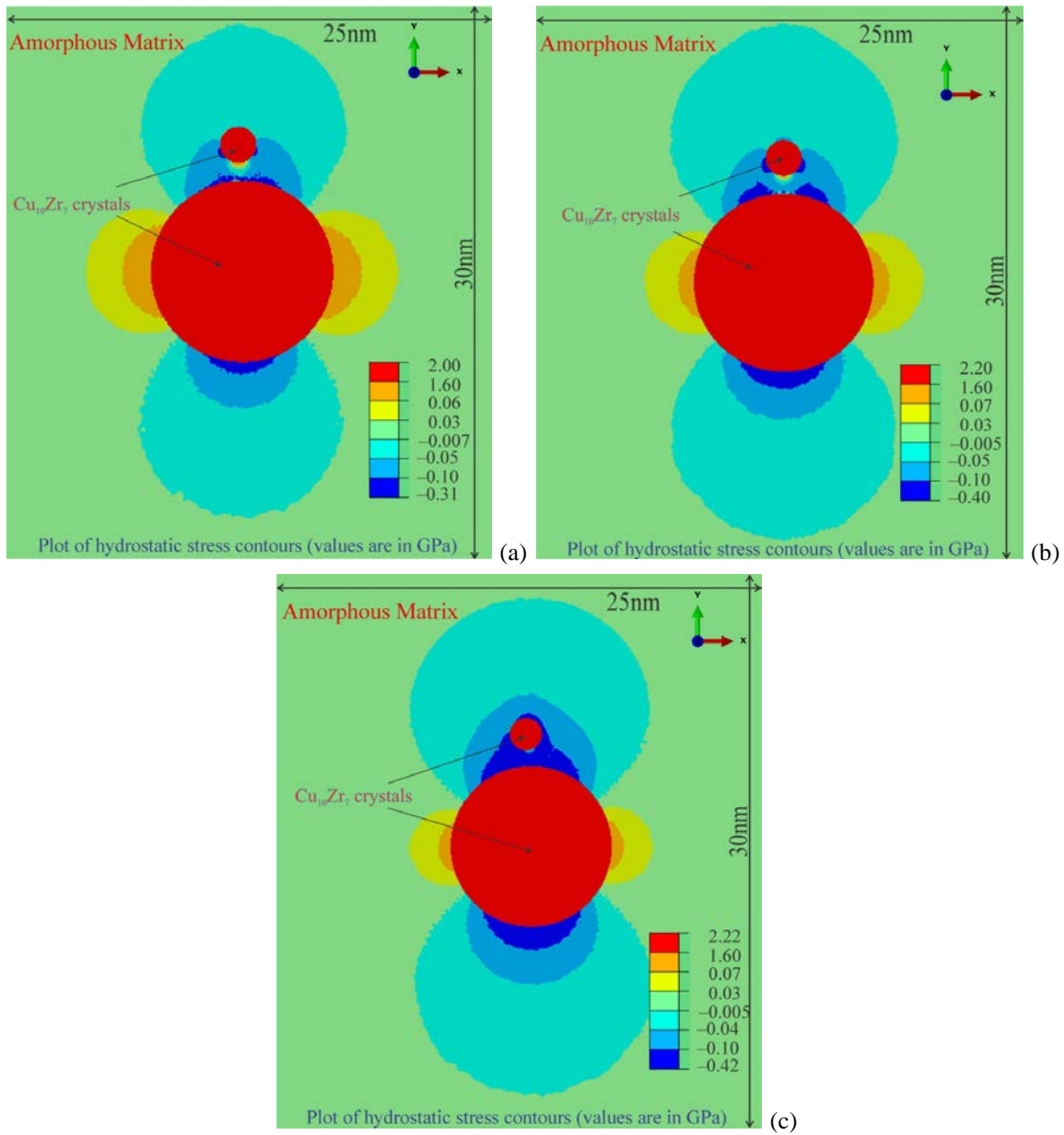

Figure 18. The effect of a change in the anisotropy (as measured by the universal anisotropy parameter ($A^U$) on the hydrostatic stress contours. (a) $A^U = 0.251$ (b) $A^U = 0.306$ (the natural value). (c) $A^U = 1.607$. The nucleus is positioned in the upper compressive lobe of the large crystal. The lower compressive lobe is practically unaffected by the presence of the nucleus and can help visualize the compressive stress contours in the absence of the nucleus. It is seen that with increasing anisotropy the magnitude of the maximum compressive (& tensile) stress increases, along with the volume of the 'highly stressed region'.

It is seen that in each of these computations (results as Table II), in spite of the broad range of the variation in the parameters, the magnitude of $r^*$ shows a change with respect to $r^*_{bulk}$. This study does not include coupled variations in the values of the parameters. The following points are noted from an analysis of the data in the table and Figure 18.

(i) As expected, an increase in the Young's modulus of either the glass or the crystal leads to an increase in $r^*_{bulk}$. However, the percentage increase is lower with an increasing modulus (in the presence of large crystal).

(ii) As evident, an increased misfit results in an increase in $r^*_{bulk}$. However, with an increasing misfit a larger decrease (in %) in $r^*$ is observed for a 'favourably' positioned nucleus.

(iii) An increased interfacial energy leads to an increase in the value of $r^*_{bulk}$ (as to be expected). The relative benefit, as measured by $\Delta r^*$ decreases (i.e. the strain energy plays a bigger role for a system with a lower interfacial energy).

(iv) An increase in the anisotropy leads to a larger region of compressive stress in the proximity of the large crystal. The magnitude of the highest compressive stress also increases. This implies that in lower symmetry systems, endowed with a higher anisotropy, are expected to exhibit a higher decrease in $r^*$ for a favorably positioned nucleus. It is to be noted that, even in cubic crystals the elastic properties are anisotropic, as the elastic constant is a fourth order tensor.

Based on the above discussion, it is reasonable to rationalize that the conclusions drawn have a wider applicability; though further investigations are required to confirm this assertion.

## 4. Summary & Conclusions

The primary conclusions are summarized in the main manuscript. In this section we consider additional points, which will help 'paint a broader picture'.

1) In the case of devitrification of a BMG, it is seen that the CNT gives a good results for the computation of $r^*$. The methodology adopted for the computation of $r^*$, in spite of the assumptions involved, captures the salient aspects of the observations− a change in the magnitude of $r^*$ due to the proximal presence of pre-existing crystals.

2) Elastic Isotropic material properties for the large crystallite and the nucleus, leads to an increase in the value of $r^*$. This implies that anisotropic elastic properties are essential to observe a decrease in the magnitude of $r^*$.

3) The shape of the nucleus plays an important role in determining the magnitude of $r^*$. The order of decreasing $r^*$ for the three shapes considered is as follows: Prolate spherioid < oblate spheroid < spherical.

4) The hydrostatic component of stress plays a key role in determining the magnitude of $r^*$. In the current example of the devitrification of a glass, the crystal is in a state of tensile stress and hence compressive hydrostatic stress in the matrix leads to a decrease in the value of $r^*$.

5) With the progress of a solid state phase transformation, the strain field of neighbouring 'grown' crystallites starts to overlap. This implies that the resultant stress field at a given point in the amorphous matrix, arises due to overlap of the stress fields arising from multiple crystallites. In regions where the accentuation of the strain field is favorable for nucleation, the decrease in the magnitude of $r^*$ can be significant.

6) Across a range of value of parameters like Young's modulus of the matrix and glass, the interfacial energy and misfit, the magnitude of $r^*$ shows a change in the proximal presence of a large crystal. This implies that the conclusions drawn in the current manuscript are expected to be applicable for a wide variety of systems.